\documentclass[12pt,a4paper]{article}
\usepackage{jheppub}
\usepackage{chemarrow}
\usepackage{amsmath,amssymb,amsfonts,mathtools} 
\usepackage{float}
\usepackage{subfig}
\usepackage{scalefnt}
\usepackage{wrapfig}
\usepackage{fancybox}
\usepackage{ifsym,yfonts}
\newcommand{\bea}{\begin{eqnarray}\displaystyle}
\newcommand{\eea}{\end{eqnarray}}
\newcommand{\nn}{\nonumber}
\newlength{\arrow}
\newcommand\es{\emptyset}
\newcommand\N{\nu_{1}}
\newcommand\NN{\nu_{2}}
\newcommand\NNN{\nu_{3}}
\newcommand\qt{Q_{\tau}}

\newcommand{\figref}[1]{Fig.~\protect\ref{#1}}

{\setlength{\fboxsep}{15pt}
\setlength{\mylength}{\linewidth}%
\addtolength{\mylength}{-2\fboxsep}%
\addtolength{\mylength}{-2\fboxrule}%
\Sbox
\minipage{\mylength}%
\setlength{\abovedisplayskip}{0pt}%
\setlength{\belowdisplayskip}{0pt}%
\equation}%
{\endequation\endminipage\endSbox
\[\fbox{\TheSbox}\]}

\begin{document}
\title{Elliptic Virasoro Conformal Blocks}
\author[a,b,c]{Amer Iqbal,}
\author[a,d]{Can Koz\c{c}az}
\author[e]{and Shing-Tung Yau}
\affiliation[a]{Center of Mathematical Sciences and Applications, Harvard University\\
1 Oxford Street, Cambridge, MA 02138, USA}
\affiliation[b]{Abdus Salam School of Mathematical Sciences\\
G. C. University, Lahore, Pakistan}
\affiliation[c]{Center for Theoretical Physics, Lahore, Pakistan}
\affiliation[d]{Jefferson Physical Laboratory, Harvard University\\
17 Oxford Street, Cambridge, MA 02138, USA}
\affiliation[e]{Department of Mathematics, Harvard University\\
1 Oxford Street, Cambridge, MA 02138, USA}

\abstract{We study certain six dimensional theories arising on $(p,q)$ brane webs living on $\mathbb{R}\times S^1$. These brane webs are dual to toric elliptically fibered Calabi-Yau threefolds. The compactification of the space on which the brane web lives leads to a deformation of the partition functions equivalent to the elliptic deformation of the Ding-Iohara algebra. We compute the elliptic version Dotsenko-Fateev integrals and show that they reproduce the instanton counting of the six dimensional theory.  }
\maketitle

\section{Introduction}
\par{The supersymmetric theories exhibit rich strong coupling phenomena in an analytically tractable setting. Over the last two decades many different approaches have been developed to construct them as well as to study their dynamics. In particular, embedding supersymmetric theories in string theory reveals mysteries of them which might be otherwise overlooked. String theory equally benefited from the progress in gauge theories to uncover various types of dualities. It provides powerful tools to compute partition functions or expectation values of observables. Independently, many exact results are obtained by localizing the path integral of theories with enough supersymmetry to matrix integral \cite{Moore:1997dj,Nekrasov:2002qd,Pestun:2007rz,Kapustin:2009kz}. These quantum field theory techniques combined with the insight endowed by string theory furnish a significantly deeper understanding in all of these different directions.    }
\par{Recently, four dimensional with ${\cal N}=2$ supersymmetry are constructed by compactification of the six dimensional $(2,0)$ theory of on a Riemann surface $C$ with punctures and are called class ${\cal S}$ theories \cite{Gaiotto:2009we}. This class is also the low energy limit of M5 branes wrapping the same Riemann surface. A surprising connection is conjectured between the conformal blocks of Liouville theory on the Riemann surface $C$ and the instanton partition function of $A_{1}$ theory on the remaining four dimension subject to $\Omega$-deformation \cite{Alday:2009aq}. This conjecture is extended to $A_{r}$ in \cite{Wyllard:2009hg}, henceforth referred to as AGTW conjecture. The correspondence depends both on the data at the punctures as well as the pants decomposition of the Riemann surface, and the specific conformal block. }
\par{The AGTW correspondence can be physically understood in terms of large $N$ duality of topological string theory \cite{Dijkgraaf:2009pc}. The class ${\cal S}$ theories can be realized as the low energy limit of a M5 brane wrapping the Riemann surface $\Sigma$, which is branched covering of $C$. This M-theory setup can be compactified by introducing a torus $T^{2}$ transverse to M5 branes. First, by shrinking along one of the circles of $T^{2}$, it becomes type IIA with NS5 branes wrapping $\Sigma$. T-duality along the second circle gives rise to type IIB string theory on a Calabi-Yau threefold whose ``base'' constitutes the Seiberg-Witten curve $\Sigma$. The topological string partition function of this Calabi-Yau threefold, in a certain limit, can be identified with the instanton partition function of the four dimensional theory. Second, at special points of the Coulomb moduli space of the gauge theory, $\Sigma$ becomes degenerate, i.e., the associated Calabi-Yau threefold becomes singular. The single M5 brane splits into multiple M5 branes at these points. The singular threefold can be blown up to obtain a non-singular threefold. The large $N$ duality connects Calabi-Yau threefold engineering the gauge theory in the B model with the blown up geometry with topological branes wrapping the two cycles. The duality identifies the closed B model to the open version, in other words, their partition functions are the same. The partition function in the presence of branes can be computed by matrix models. The final ingredient to understand the AGTW conjecture in this context is the Dotsenko-Fateev representation of the conformal blocks \cite{Dotsenko:1984nm,Dotsenko:1984ad,Mironov:2010zs,Mironov:2010su} which is equal to the matrix model of the open amplitude. Although this approach gives an excellent physical intuition, it is still at the level of a conjecture since the large $N$ duality is a conjecture. It is also restricted to the self-dual $\Omega$-background, and can not cover the generic background. The M-theory offers a formulation for refined topological string theory on non-compact Calabi-Yau threefolds with additional $U(1)$ isometry. Further details can be found in the detailed review \cite{Aganagic:2014kja}.   }
\par{The AGTW conjecture has been lifted to a conjecture of one dimension higher theory on the gauge theory side and to the $q$-deformation on the conformal theory side. Instead of considering the instanton partition function of ${\cal N}=2$ in four dimensions, one can study the ${\cal N}=1$ five dimensional theory compactified on a circle. Initially, the instanton partition function of pure $SU(2)$ theory was checked up to 9 instantons to match the norm of the Gaiotto-like state \cite{Gaiotto:2009ma} of the $q$-deformed Virasoro algebra \cite{Awata:2009ur}. Later, the $q$-deformed version of the $\beta$-ensemble was proposed and shown to coincide with the instanton partition function of five dimensional of $SU(N)$ theory with $N_{f}=2N$ flavors \cite{Awata:2010yy}.     }
\par{The $q$-deformed AGTW correspondence is enhanced to a triality when considered within the context of topological string theory \cite{Aganagic:2013tta,Aganagic:2014oia,Aganagic:2015cta}. As we will elaborate further details of this construction, they showed that the 5d instanton partition function of the ${\cal N}=1$ theory can be studied at a particular point of its moduli space and the instanton partition function reduces to their 3d vortex partition function. This is reminiscent of the duality analyzed as a 2d/4d correspondence in \cite{Dorey:2011pa,Chen:2011sj}, which is related to the brane construction of the vortices \cite{Hanany:2003hp,Hanany:2004ea}. The 3d vortex partition function can be computed using localization, similar to the 2d case in \cite{Shadchin:2006yz}, and it is identical to the matrix integrals appearing in Dotsenko-Fateev representation of the conformal blocks. The very non-trivial fact introduced by lifting the AGTW duality one dimension higher is based on the so-called fiber-base duality in topological string theory. The ${\cal N}=1$ theories can be geometrically engineered by compactifying M-theory on a Calabi-Yau threefold \cite{Katz:1996th}. For the $A_{N}$ type theories, the threefold is a non-compact toric Calabi-Yau which is an $A_{N}$ fibration over a collection of ${\mathbb P}^{1}$'s whose intersection can be mapped to the Cartan matrix of, say $A_{M}$. In \cite{Katz:1997eq}, a duality is observed that exchanges the base $A_{M}$ with the fiber $A_{N}$. The topological string theory partition function on this toric Calabi-Yau threefold is obviously invariant under this exchange, but the gauge theory interpretation in the transverse 5d are quite different. We will call such dual gauge theories fiber-base duals. The simplest example of such a pair is the $SU(2)\times SU(2)$ theory with a bi-fundamental matter multiplet and an $SU(3)$ theory with two hypermultiplets in the fundamental and anti-fundamental representation. The same topological string partition function gives rise to the instanton partition function of two different gauge theories in 5d depending whether it expanded in terms of the base or fiber K\"{a}hler classes. However, the instanton partition functions of the fiber-base duals in 4d are very different from each other. The origin of the difference is the dissimilar scaling, in the 4d field theory limit, on the fiber and base K\"{a}hler classes. The unifying nature of the lift from 4d to 5d is the crucial ingredient. In \cite{Aganagic:2013tta}, the $q$-deformed Virasoro conformal block is shown to be identical to the instanton partition function of the gauge theory whose fiber-base dual is the 5d lift of the gauge theory appearing in the AGTW duality. The invariance provides a \textit{physics proof} of the duality \footnote{The invariance of the instanton expansions for fiber-base duals can be shown order by order but a general mathematical proof is still lacking. For more sophisticated approaches, we refer \\ the reader to \cite{Zenkevich:2014lca,Morozov:2015xya}.}. In that work, the momenta carried by the vertex operators are quantized and are given by the vortex charges (giving a physical explanation as to why the momenta in Dotsenko-Fateev integrals are quantized). However, later in \cite{Aganagic:2014kja}, the large $N$ limit is suggested to make the equivalence for arbitrary momenta.   }
\par{In this paper, we are extending the relation between the instanton partition function of the 5d theory and the $q$-deformed Liouville conformal blocks. In \cite{Aganagic:2003db}, a partial compactification of $(p,q)$ brane webs dual to certain toric Calabi-Yau threefolds were introduced and this construction was argued to be related to the 6d lift of the 5d theory. The Calabi-Yau threefolds dual to these partially or completely compact brane webs, living in $\mathbb{R}\times S^{1}$ or $T^2$, are elliptically fibered. We show that this partial compactification of brane webs, related to elliptic Calabi-Yau threefolds, corresponds algebraically to elliptic deformation proposed in \cite{saito}. The motivation for the prescription is to formulate the elliptic version of Ding-Iohara algebra. This algebra has already appeared in \cite{Awata:2011dc} to study $q$-deformation of AGTW conjecture. The vertex operators and the screening charges appearing in $q$-deformed Dotsenko-Fateev integrals posses a free boson representation. They can be subjected to the elliptic deformation in \cite{saito}. We will call the matrix integral of these elliptically deformed operators \textit{elliptic Virasoro conformal blocks}. In the same flavor as in \cite{Aganagic:2013tta}, we show the match between the 6d instanton partition functions and elliptic conformal blocks. This type of connection was argued from M-theory perspective in \cite{Tan:2013xba}.  }
\par{This paper is organized as follows. In section 2, we will present the connection between the $q$-deformed Liouville and 5d instanton partition functions. In this section we also introduce the mathematical details of these theories which we will need later. In section 3, we review the elliptic deformation and it geometric interpretation as partial compactification of brane webs dual to toric Calabi-Yau threefolds. Section 4 is reserved to the elliptic deformation of vertex operators and screening charges, the building blocks of the Dotsenko-Fateev integrals. In section 5, we work out three examples and make our general proposal for the elliptic extension of 5d instanton/$q$-deformed conformal block correspondence. }
\\
\par{We understand that related results have been obtained independently by Fabrizio Nieri. We thank him for coordinating submission of his work \cite{nieri} with us.}
\section{Gauge/Liouville Triality}

\par{In this section we would like to review the Gauge/Liouville triality, introduced for the first time in \cite{Aganagic:2013tta} for ${\cal N}=2$ $A_{1}$ theories and extended later to other gauge groups \cite{Aganagic:2014oia,Aganagic:2015cta}. The triality is motivated by the earlier observation of the duality between a 4d ${\cal N}=2$ gauge theory and a 2d ${\cal N}=(2,2)$ theory on its vortices \cite{Dorey:2011pa,Chen:2011sj}. The coincidence of the BPS spectra of these theories were first hints for such a duality \cite{Dorey:1998yh,Dorey:1999zk}.  }
\par{The 4d ${\cal N}=2$ theory has the gauge group $SU(L)$ with $L$ hypermultiplets in the fundamental representation and $L$ in the anti-fundamental representation subject to reduced $\Omega$-background, i.e. one of the deformation parameters $\epsilon_{1}$ is set to zero. Along the non-zero deformation parameter $\epsilon_{2}$, ${\cal N}=(2,2)$ supersymmetry is preserved. The resulting 2d effective theory has a twisted superpotential. The supersymmetric vacua of this effective theory constitute an $L$ dimensional lattice and are determined by the F-term equation }
\begin{align}
\vec{a}={\vec m}_{F}-\vec{n}\epsilon_{2},\,\,\mbox{for}\,\qquad\qquad \vec{n}=(n_{1},\mathellipsis,n_{L})\in\mathbb{Z}^{L},
\end{align}
where the vector $\vec{a}$ label the Coloumb branch parameters and ${\vec m}_{F}$ the masses of the hypermultiplets in the fundamental representation. The 2d part with the $\Omega$-deformation of the 4d spacetime has the topology of a cigar. The 2d effective theory corresponds to the theory on its vortices with vortex charges given by $n_{i}$. In the absence of the $\Omega$-deformation, the vortex charge can not be turned on without introducing a surface operator insertion \cite{Gukov:2008ve}. However, in the presence of non-zero $\epsilon_{2}$ one can turn them on and they effectively shift the Coulomb branch moduli. At the level of F-terms, once the $\Omega$-background is switched on, turning on the vortex flux is the same as shifting the Coulomb branch parameters by an amount proportional to flux. 
\par{The 2d theory of the duality has ${\cal N}=(2,2)$ supersymmetry and $U(N)$ gauge group. This is the theory of the vortices themselves at the root of the Higgs branch of the 4d theory, even if the $\Omega$-deformation is off. It has $L$ chiral multiplets in the fundamental and anti-fundamental representation, as well as a single chiral multiplet in the adjoint representation with the mass $\epsilon_{2}$. The FI parameters form the complex marginal coupling with the vacuum angle. The claim of the duality is that the supersymmetric vacua of these two theories described are in one-to-one correspondence once the parameters of both theories are appropriate tuned. Moreover, their twisted superpotentials are the same on-shell. }
\par{The above duality can be lifted to one dimension higher. The 4d theory can be replaced by a circle fibration of the reduced $\Omega$-background, with ${\cal N}=1$ supersymmetry. Similarly, the dual 2d theory can be lifted to a theory on a 3d space time of the same type circular fibration, with the same radius. One needs to take into account an infinite tower of Kaluza-Klein modes. The duality is expected to survive the lift. One part of the triality in \cite{Aganagic:2013tta} explicitly shows that the instanton partition function of the full 5d $\Omega$-background is the same as the one of the 3d theory.}
\par{The last part of the triality depends on the observation that the partition function of the 3d theory is identical to the $q$-deformed Liouville conformal blocks, hence is identical to the 5d instanton partition function. The path integral of the 3d gauge theory can be reduced to a matrix integral by employing the supersymmetric localization. As we will discuss shortly, the conformal blocks of the usual and $q$-deformed Liouville theory can be written as matrix integrals, and the resulting matrix integrals of the 3d theory and the $q$-deformed conformal blocks are identical under a suitable dictionary of variables of two theories. Few remarks are in order to clarify some subtle points. First, the equivalence between the partition functions of the 3d theory and the 5d theory is valid at the special points in the moduli space where the Higgs branch of the theory meets its Coulomb branch, and turning on vortex fluxes. Any other point in the Coulomb branch can be probed by taking the large vortex flux limit such that $n_{i}$'s go to infinity while $\epsilon_{2}$ is sent to zero, keeping their product to any finite value. Second, the duality is expected to reduce to aforementioned 4d-2d duality when the radius of the circle is taken to zero, decoupling all the Kaluza-Klein tower. The $q$-deformed Liouville theory becomes the usual one. Finally, as one might expect, the 4d instanton partition function identical to the Liouville conformal blocks, but up to the so-called fiber-base duality or spectral duality. The reason behind this duality is simply the fact that the taking the radius to zero is not a unique limit. There is a variety of ways to take this limit, and different choices lead to different 4d theories. For instance, the AGT conjecture states that the instanton partition function of the $SU(2)\times SU(2)$ theory with two flavors under each gauge group and one bi-fundamental hypermultiplet is the same as the Liouville conformal block of the five point function on the sphere. However, according to the Gauge/Liouville triality the five point $q$-deformed Liouville conformal is identical to the instanton partition function of $SU(3)$ theory with six flavors. Indeed, it is known fact in geometric engineering that these two theories share the same topological string theory partition function. In other words, expanding the topological string partition function in different K\"{a}hler parameters give rise to the instanton partition function of these dual theories in 5d. It is worth to mention that, up to the fiber-base duality, the Gauge/Liouville triality is a direct physics proof of the AGT conjecture.      }

\subsection{Liouville Theory Conformal Blocks}

\par{In this paper, our focus is the connection between the instanton partition function of the 6d theory and the \textit{elliptic conformal blocks}. Therefore, we prefer to review the free field representation of the Liouville conformal blocks due to Dotsenko and Fateev. The action of the Liouville conformal field theory can be written in terms of a boson as  }
\begin{align}
S=\int dzd{\bar z}\,\sqrt{g}\left[g^{z{\bar z}}\partial_{z}\phi\partial_{{\bar z}} \phi+Q\phi R+e^{2b\phi}\right],
\end{align}
where $b$ is the coupling constant, $Q=b+b^{-1}$ denotes the background charge. The central charge of the theory can be expressed in terms of the background charge $c=1+6Q^{2}$. We can insert primary fields with momenta $\alpha$ at points $z$ whose vertex operator take the following form,
\begin{align}
V_{\alpha}(z)=:\exp\left(-\frac{\alpha}{b}\phi(z) \right):,
\end{align}
where $:\mathellipsis:$ stands for the normal ordering. The conformal block with primary operator insertions has a Coulomb gas representation
\begin{align}\label{cb1}
{\cal B}_{\alpha_{0},\mathellipsis,\alpha_{M+1}}(z_{1},\mathellipsis,z_{M})=\oint dy_{1}\mathellipsis\oint dy_{n}\,\langle V_{\alpha_{0}}(0)\mathellipsis V_{\alpha_{M}}(z_{M})V_{\alpha_{M+1}}(\infty)S(y_{1})\mathellipsis S(y_{n})\rangle,
\end{align}
where we have fixed two insertions to be at $0$ and $\infty$ although we could have fixed yet another insertion point by the conformal symmetry. In addition to the primary fields, we inserted screening charges 
\begin{align}
S(y)=:e^{2b\phi(y)}:,
\end{align}
and their insertions come from treating the Liouville potential as perturbation and bringing powers of them down. Representation of the conformal block in the form of Eq.(\ref{cb1}) requires the balancing condition for the $U(1)$ charges of the vertex operators,
\begin{align}
\frac{\alpha_{\infty}}{b}+\sum_{i=0}^{M}\frac{\alpha_{i}}{b}=2bn+RQ.
\end{align}
\begin{figure}[ht]
  \centering
  \includegraphics[width=2.5in]{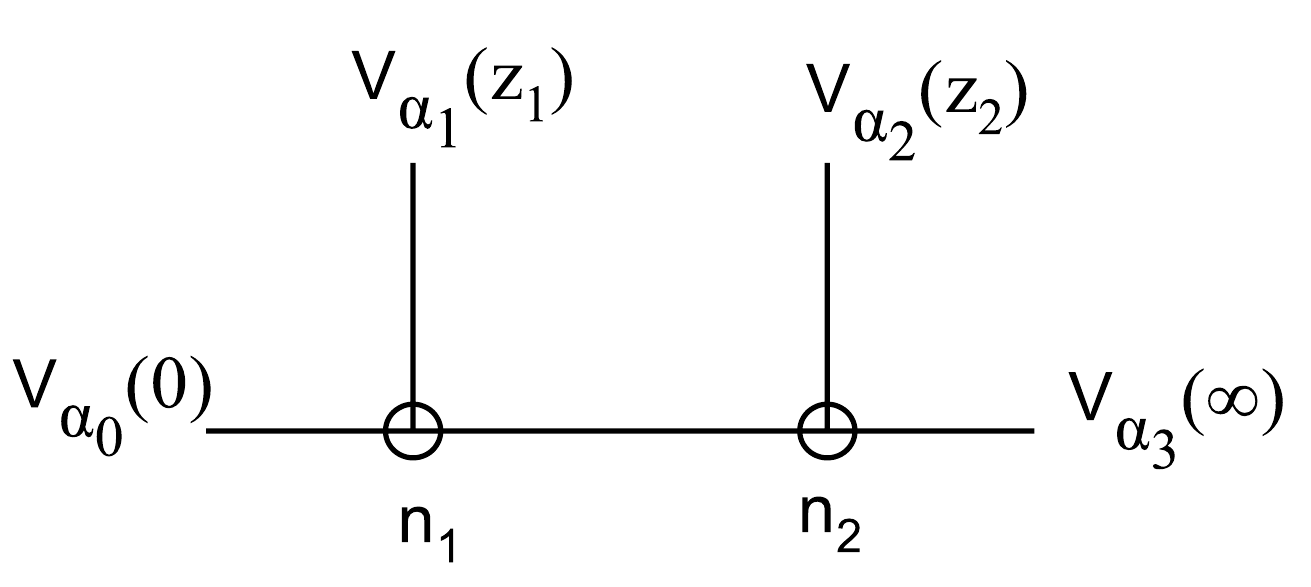}\\
  \caption{The 4-point conformal block}\label{block}
\end{figure}
Using the free mode expansion of the chiral field $\phi(z)$, it is easy to write an integral expression for the conformal blocks. Let us write the chiral field in terms of annihilation and creation operators
\begin{align}
b\phi(z)=\phi_{0}+h_{0}\log z+\sum_{n\neq0}\textswab{h}_{n}\frac{z^{-n}}{n},
\end{align}
where $\phi_{0}$ is a constant and the modes $\textswab{h}_{k}$ obey the usual Heisenberg algebra,
\begin{align}
[\textswab{h}_{n},\textswab{h}_{m}]=\frac{-b^{2}}{2}n\,\delta_{n+m,0},\qquad\qquad\mbox{for}\,\,n,m\in\mathbb{Z}.
\end{align}
The two point functions of between vertex operators and screening charges can be easily computed in terms of the mode expansions,
\begin{align}
\langle V_{\alpha}(z)V_{\alpha'}(z')\rangle&=(z-z')^{\frac{-\alpha\alpha'}{2b^{2}}},\\\nonumber
\langle V_{\alpha}(z)S(y)\rangle&=(z-y)^{\alpha},\\\nonumber
\langle S(y)S(y')\rangle&=(y-y')^{-2b^{2}},
\end{align}
and the Coulomb gas representation of the conformal blocks takes the following form
\begin{align}
{\cal B}_{\alpha_{0},\mathellipsis,\alpha_{M+1}}(z_{1},\mathellipsis,z_{M})=\frac{C}{\prod_{a=1}^{M}n_{a}!}\oint_{\vec{\gamma}} d^{n}y\, \Delta_{\beta}^{2}(y)\,\prod_{a=0}^{M}V_{a}(y),
\end{align}
where $\Delta_{\beta}^{2}(x)$ is the $\beta$-deformed Vandermonde with $\beta=-b^{2}$, and is given by
\begin{align}
\Delta_{\beta}^{2}(y)=\prod_{1\leq i<j\leq N}(y_{i}-y_{j})^{2\beta},
\end{align}
and the potential $V_{\alpha}(y)$ reads
\begin{align}
V_{a}(y;z)=\prod_{i=1}^{N}(y_{i}-z_{a})^{\alpha_{a}}.
\end{align}
\begin{figure}[ht]
  \centering
  \includegraphics[width=2in]{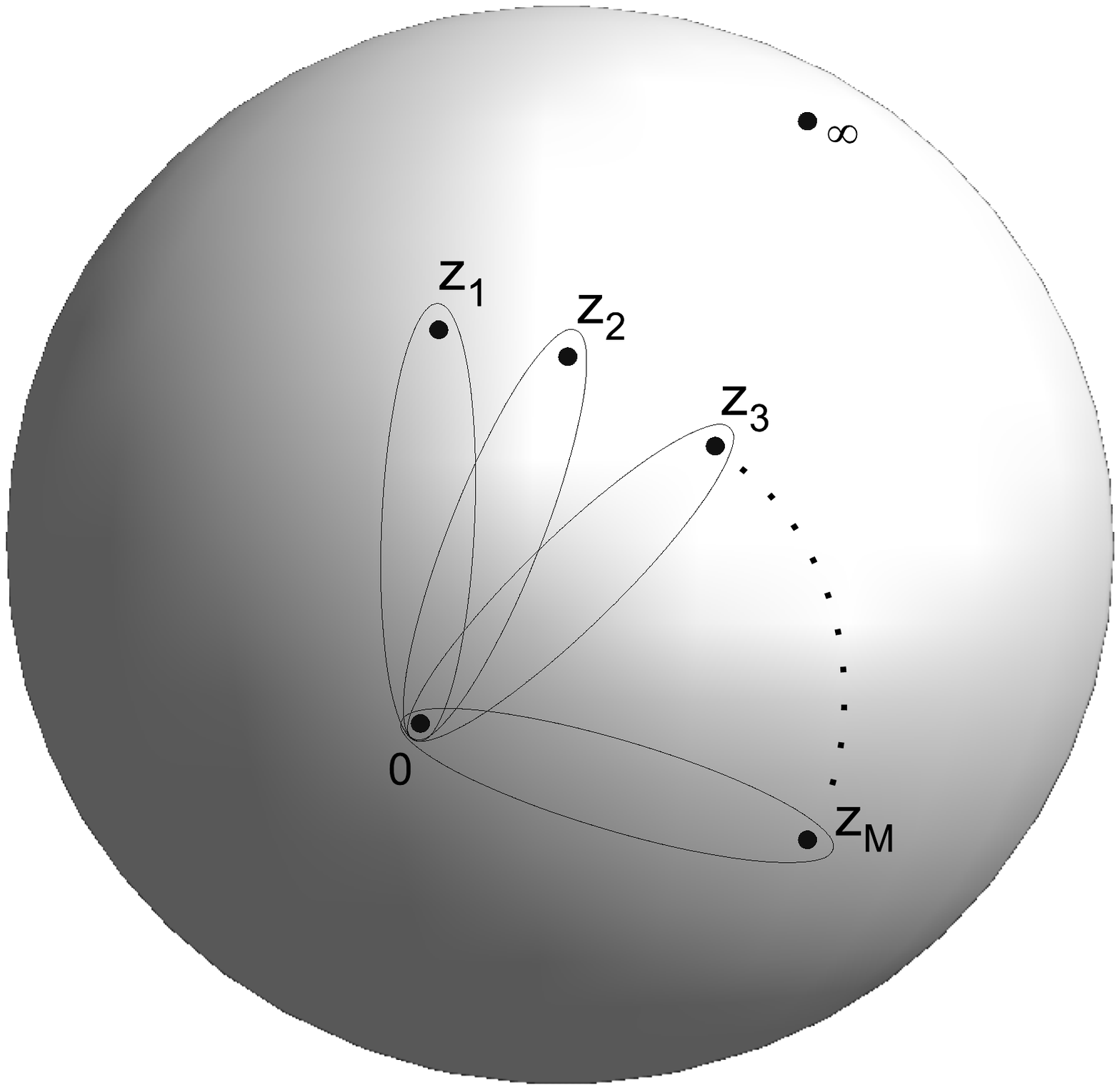}\\
  \caption{Each integration contours have been shown to encircle the origin and the insertion point of one of the vertex operators $V_{z}(z_{a})$ for $a=1,\mathellipsis,M$.}\label{contour}
\end{figure}
Note that $\sum_{a=1}^{M}n_{a}=n$. The above integral is ambiguous as it stands unless the contours $\vec{\gamma}$ are defined. Motivated by the AGT conjecture, the contours are chosen and checked in \cite{Mironov:2010zs}. As already mimicked by the pre-factor in front of the integral, the $n$ integrals over the screening charges are split into $M$ groups of $n_{a}$ variables. The contour for the integration variables from the $a$-th group encircles a segment $\gamma_{a}=[0,z_{a}]$. The factor $C$ includes factors which are irrelevant for our purposes, 
\begin{align}
C=\prod_{1\leq a<b\leq M}(z_{a}-z_{b})^{\frac{\alpha_{a}\alpha_{b}}{2\beta}}.
\end{align}

\subsection{$q$-deformed Liouville Theory Conformal Blocks}

\par{The $q$-deformation of the Liouville theory was first constructed using the quantum group techniques \cite{Shiraishi1,Shiraishi2}. The vertex operators and screening charges are modified two complex variables, $q$ and $t=q^{\beta}$ for a generic complex parameter $\beta$. The operators and the algebra are adjusted in a way that the limit $q\rightarrow1$ reduces the theory to the usual Liouville theory. For completeness, we briefly review here the original construction, although we use a slightly different construction. The $q$-deformed vertex operators take the following form,}

\begin{align}
V_{\alpha}(z)=\,:\exp\left(\frac{\alpha}{\beta}q+\frac{\alpha}{\beta}p\log z +\sum_{n\neq 0}\frac{1}{n}\frac{1-q^{\alpha n}}{1-t^{n}}\textswab{h}_{n} z^{n}\right):\,,
\end{align}
where the modes satisfy a $q$-deformed Heisenberg algebra,
\begin{align}
[\textswab{h}_{n},\textswab{h}_{m}]=\frac{1}{1+(q/t)^{n}}\frac{1-t^{n}}{1-q^{n}}n\delta_{n+m,0}.
\end{align}
The screening charges are equally modified and are given by
\begin{align}
S(y)=:\exp\left(2q+2p\log z+\sum_{n\neq0}\frac{1+(q/t)^{n}}{n} \textswab{h}_{n}z^{n}\right):.
\end{align}
The Dotsenko-Fateev integral can be easily written after computing similar two point functions as in the undeformed case. It has the same form except the Vandermonde becomes the $q$-deformed one
\begin{align}
\Delta_{q,t}^{2}(y)=\prod_{1\leq i\neq j\leq N}\frac{\varphi(y_{i}/y_{j})}{\varphi(t\,y_{i}/y_{j})},
\end{align}
where $\varphi(z)$ is the quantum dilogarithm
\begin{align}
\varphi(z)=\prod_{n=0}^{\infty}(1-z\,q^{n}).
\end{align}
Similarly, the potentials become functions of the quantum dilogarithm function
\begin{align}
V_{a}(y;z_{a})=\prod_{i=1}^{N}\frac{\varphi(q^{\alpha_{a}}z_{a}/y_{i})}{\varphi(z_{a}/y_{i})}.
\end{align}
The factor in front of the integral still remains but also $q$-deformed and we will ignore them since we want to focus on the instanton partition function on the gauge theory side. It turns out the contour remain the same; encircling again the given segments. 

\subsection{5d Instanton Partition Function}

\par{In this section, we review the instanton partition function of the 5d ${\cal N}=1$ $SU(N)$ gauge theory with $N_{f}=2N$ on $\mathbb{R}^{4}\times S^{1}$ subject to $\Omega$-background,}
\begin{align}
M_{q,t}=\mathbb{C}^{2}_{q,t^{-1}}\times S^{1},
\end{align}
where the product denotes a fibration; as we go around the $S^{1}$ there is a $U(1)\times U(1)$ action on $\mathbb{C}^2$ given by
\bea
(z_{1},z_{2})\mapsto (e^{i\epsilon_{1}}z_{1},e^{i\epsilon_{2}}z_{2})\,,
\eea
with $(e^{i\epsilon_{1}},e^{i\epsilon_{2}})=(q,t^{-1})$. There is an accompanying 5d $U(1)_{R}$ twist to preserve supersymmetry. The instanton partition function of the 5d theory on this background can be computed using the supersymmetric localization techniques which reduces it to an equivariant integral over the instanton moduli spaces. If we denote by ${\cal M}(N,k)$ the $SU(N)$ instanton moduli space of charge $k$ then the instanton partition function ${\cal Z}$ is given by,
\bea\label{int1}
{\cal Z}=\sum_{k\geq 0}\,\Lambda^{k}\,\int_{{\cal M}(N,k)}\,\prod_{i}\frac{x_{i}}{1-e^{-x_{i}}}\prod_{a=1}^{N_{f}}(1-y_{a}e^{-\tilde{x}_{i}})\,.
\eea
where $x_{i}$ are the roots of the Chern polynomial of the tangent bundle to ${\cal M}(N,k)$ and $\tilde{x}_{i}$ are the roots of the Chern polynomial of the bundle corresponding the matter content. In the case we are interested, the case of fundamental hypermultiplets, the matter fields are sections of a tautological bundle $V$ on the instanton moduli space. Recall that ${\cal M}(N,k)$ is a hyper-K\"{a}hler quotient with $\text{dim}_{\mathbb{R}}\big({\cal M}(N,k)\big) = 4Nk$ and is defined through the following quotient:
\bea
M_{N,k}/U(k)
\eea
where 
\bea\nn
M_{N,k}=\{(B_{1},B_{2},i,j)\,|\,[B_1,B_2]  + ij =0\,,\,[B_1 , B_1^{\dagger}] + [B_2, B_2^{\dagger}] - ii^{\dagger} - jj^{\dagger} = \zeta Id\,\}\,.
\eea
and $B_{1,2}\in \text{End}(V), i \in \text{Hom}(V, W),  j \in \text{Hom}(W, V)$ such that $V$ is a $k$ dimensional vector space and $W$ is an $N$ dimensional vector space. The $U(k)$ action is defined as
\bea g(B_1, B_2, i, j) = (g B_1  g^{-1}, g B_2 g^{-1}, g i, j g^{-1}). \eea ${\cal M}(N,k)$ has a $U(1)^N \times U(1)_{\epsilon_1} \times U(1)_{\epsilon_2}$ action defined on it ($e=\mbox{diag}(e_1,e_2,\mathellipsis,e_N)$):
\bea \label{action}
(B_1,B_2,i,j)\mapsto (e^{i\epsilon_1} B_{1},e^{i\epsilon_2} B_{2},i\,e^{-1}, e\,j).\eea 
This $T^{N+2}$ action on ${\cal M}(N,k)$ has fixed points which are in one-to-one correspondence with $N$-tuples of partitions $(\nu_1, \mathellipsis, \nu_N)$ such that $|\nu_1| + \mathellipsis+ |\nu_N| = k$.  The weights of this action on $T_{[\vec{Y}]} M(N,k)$ at the fixed point labelled by $(\nu_1, \mathellipsis, \nu_N)$, denoted by $w_{r}$, are given by \cite{Flume:2002az,Bruzzo:2002xf},
\bea\label{tangentweights}\nn
\sum_{r}e^{w_{r}}=\sum_{\alpha,\beta=1}^{N}e_{\beta}e_{\alpha}^{-1} \Big(\sum_{(i,j)\in \nu_{\alpha}}q^{-(\nu_{\beta,j}^{t}-i)}t^{-(\nu_{\alpha,i}-j+1)}+\sum_{(i,j)\in \nu_{\beta}}q^{\nu_{\alpha,j}^{t}-i+1}t^{\nu_{\beta,i}-j}\Big).
\eea
We denote by ${\mathbb V}$ the vector bundle over ${\cal M}(N,k)$ whose fiber over a point $(B_{1},B_{2},i,j)$ is given by the vector space $V$. This equivariant bundle has weight decomposition \cite{nakajima1,Li:2004ef},
\bea
{\mathbb V}&=&\oplus_{\alpha}\,V_{\alpha}e_{\alpha}\,,
\eea
such that on $V_{\alpha}$ the weights of the equivariant action, denoted by $u^{\alpha}_{p}$, are given by,
\bea
\sum_{p}e^{u^{\alpha}_{p}}&=&\sum_{(i,j)\in \nu_{\alpha}}q^{j-1}\,t^{-i+1}\,.
\eea
With $N_{f}$ fundamental hypermultiplets therefore the equivariant bundle will be,
\bea
V\otimes \mathbb{C}^{N_f}\,.
\eea
There is global symmetry $U(N_f)$ acting on $\mathbb{C}^{N_f}$. The parameters giving the Cartan $U(1)^{N_{f}}\subset U(N_f)$ are related to the masses of the fundamental hypermultiplets. Including this $U(1)$ action on the $a-th$ copy of $V$ the weights of the equivariant action are given by
\bea
\sum_{p}e^{u^{\alpha}_{p}}=y_{a}\sum_{(i,j)\in \nu_{\alpha}}q^{j-1}\,t^{-i+1}\,.
\eea

The integration in Eq.(\ref{int1}) is carried out equivariantly and receives contributions only from fixed points of the instanton moduli spaces which are labelled by Young diagrams.  The partition function becomes an explicit sum over all possible Young diagrams. The summand consists of factors coming from vector multiplet and different types of matter multiplets
\begin{align}
{\cal Z}=\sum_{\vec{\nu}}\Lambda^{|\vec{\nu}|}\, T_{\vec{\nu}}\,\Big(z^{vector}_{\vec{\nu}}\times z^{matter}_{\vec{\nu}}\Big)
\end{align}
where $\vec{\nu}=(\nu_{1},\nu_{2},\mathellipsis,\nu_{N})$ is the $N$-tuple of Yound diagrams and $T_{\vec{\nu}}=\prod_{a=1}^{N}(-1)^{|\nu_{a}|}q^{\Arrowvert\nu_{a}\Arrowvert^{2}/2}t^{-\Arrowvert\nu_{a}^{t}\Arrowvert^{2}/2}$. The vector multiplet contributions can be expressed in terms of Nekrasov function as
\begin{align}
z^{vector}_{\vec{\nu}}=\prod_{r}(1-e^{w_{r}})^{-1}=\prod_{1\leq a,b\leq N}\frac{1}{N_{\nu_{a}\nu_{b}}(e_{a}/e_{b})},
\end{align}
where the Nekrasov function has different equivalent forms, one of which is 
\begin{align}
N_{\nu\mu}(Q)=\prod_{i,j=1}^{\infty}\frac{\varphi(Q\,q^{\nu_{i}-\mu_{j}}t^{j-i+1})}{\varphi(Q\,q^{\nu_{i}-\mu_{j}}t^{j-i})}\frac{\varphi(Q\,t^{j-i})}{\varphi(Q\,t^{j-i+1})}.
\end{align}
Similarly, the contributions of $N_{f}$ hypermultiplets in the fundamental and anti-fundamental representations are given by
\begin{align}
z^{fund}_{\vec{\nu}}&= \prod_{a=1}^{N_{f}}\prod_{\alpha}\prod_{p}(1-e^{u^{\alpha}_{p}})=\prod_{1\leq a\leq N_{f}}\prod_{1\leq b\leq N}N_{\emptyset\nu_{b}}(vf^{+}_{a}/e_{b}),
\end{align}
respectively such that $f_{a}^{+}=y_{a}$.

\subsubsection{Partition Function from Geometry}
\begin{figure}[ht]
  \centering
  \includegraphics[width=3.5in]{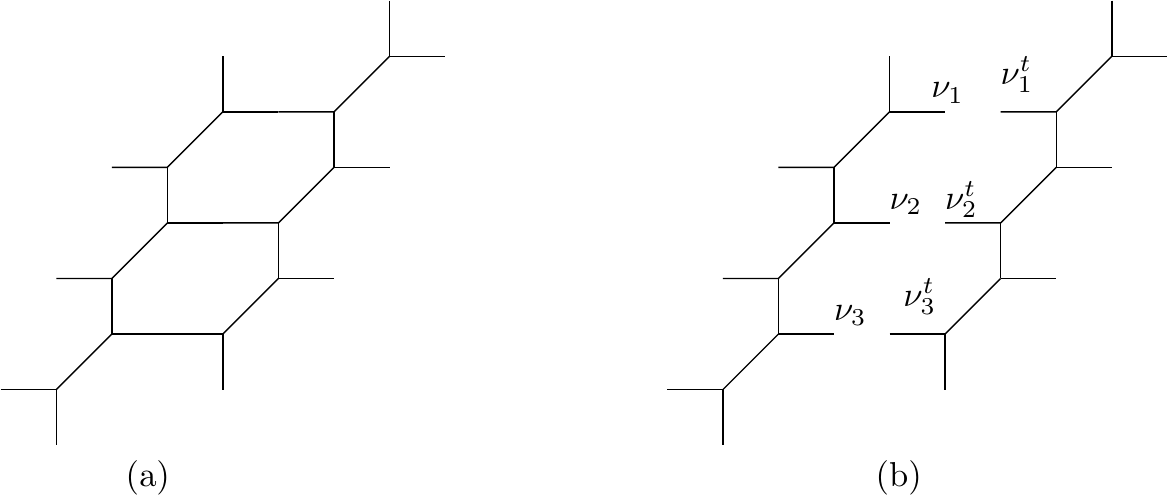}\\
  \caption{(a) The web dual to toric Calabi-Yau threefold $X_{3,6}$. (b) The topological string partition function can be computed using the refined topological vertex, first, dividing the geometry into two halves and computing the open amplitudes on both sides. Then we glue them along  the instanton direction.}\label{su3}
\end{figure}

The five dimensional ${\cal N}=1$ $SU(N)$ gauge theory with $N_{f}$ fundamental hypermultiplets can be geometrically engineered using M-theory on a Calabi-Yau threefold. The Calabi-Yau threefold in this case is resolved $A_{N-1}$ singularity blown up at $N_{f}$ points fibered over a $\mathbb{P}^1$ which we will denote by $X_{N,N_{f}}$. The partition function of the five dimensional gauge theory is given by the topological string partition function of $X_{N,N_{f}}$. For $N_{f}\leq 2N$ $X_{N,N_{f}}$ is a toric Calabi-Yau threefold. The case $N_{f}=2N$ that will be of importance for us later has a dual web diagram shown in \figref{su3}(a).

The refined topological string partition function of this geometry can be determined using the refined topological vertex formalism \cite{Iqbal:2007ii} and is given by
\bea
{\cal Z}=\sum_{\vec{\nu}}\Lambda^{|\vec{\nu}|}\,W^{vector}_{\vec{\nu}}\,W^{fund}_{\vec{\nu}}
\eea
 where
\bea\nonumber
W^{vec}_{\vec{\nu}}&=&\frac{\prod_{a=1}^{N}\Big(q^{\frac{\Arrowvert\nu_{a}\Arrowvert^2}{2}}\,t^{\frac{\Arrowvert\nu^{t}_{a}\Arrowvert^2}{2}}\,Z_{\nu_{a}}(t,q)Z_{\nu^{t}_{a}}(q,t)\Big)}
{\prod_{i,j}\prod_{1\leq a<b\leq N}\Big(1-Q_{ab}\,t^{-\nu^{t}_{a,i}+j}\,q^{-\nu_{b,j}+i-1}\Big)\Big(1-Q_{ab}\,t^{-\nu^{t}_{a,i}+j-1}\,q^{-\nu_{b,j}+i}\Big)}\\\nonumber
&=&\frac{T_{\vec{\nu}}}{\prod_{a,b=1}^{N}N_{\nu_{a}\nu_{b}}(Q_{ab})}\,,
 \eea
and
\bea\nn
W^{fund}_{\vec{\nu}}&=&\prod_{i,j}\prod_{1\leq a\leq b\leq N}\Big(1-Q_{ab}Q_{m_{b}}\,t^{-\nu^{t}_{a,i}+j-\frac{1}{2}}\,q^{i-\frac{1}{2}}\Big)\Big(1-Q_{ab}Q_{m_{a}}\,t^{j-\frac{1}{2}}\,q^{-\nu_{b,j}+i-\frac{1}{2}}\Big)\\\nonumber
&\times&\prod_{a<b}\Big(1-Q_{ab}Q_{m_{a}}^{-1}\,t^{j-\frac{1}{2}}q^{-\nu_{b,j}+i-\frac{1}{2}}\Big)\Big(1-Q_{ab}Q_{m_{b}}^{-1}\,t^{-\nu^{t}_{a,i}+j-\frac{1}{2}}\,q^{i-\frac{1}{2}}\Big)\,.
\eea
In the above equations,
\bea
Q_{m_{a}}=f_{a}^{+}=e^{2\pi i m_{a}}\,,\,\,\,\,\,Q_{ab}=e_{b}\,e_{a}^{-1}\,.
\eea
$W^{vec}_{\vec{\nu}}$ contains the contribution of all torus invariants holomorphic curves with local geometry ${\cal O}(-2)\oplus {\cal O}(0)\mapsto \mathbb{P}^1$ and $W^{fund}_{\vec{\nu}}$ contains the contribution of all holomorphic curves with local geometry ${\cal O}(-1)\oplus {\cal O}(-1)\mapsto \mathbb{P}^1$.

\subsection{Truncation of the Instanton Partition Function}

\par{As we have explained before, we will probe the 5d/6d gauge theory at a very special point in their moduli space; where the Coulomb branch meets the Higgs branch. This point is reached if we tune the masses of the hypermultiplets in the fundamental representation to be equal to the Coulomb branch parameters up to an integer multiple of the $\epsilon_{2}$: }
\begin{align}
e_{a}=\frac{t^{n_{a}}}{v}f^{+}_{a}
\end{align}
Remember that the contribution from the hypermultiplets in the fundamental representation includes factors $N_{\emptyset\nu_{a}}(vf^{+}_{a}/e_{a})$ for each multiplet, $a=1,\mathellipsis,M$. This identification imposes a restriction on the length of each Young diagram $\nu_{a}$ that we are summing over. It is very easy to see if one considers the following representation of the Nekrasov function,
\begin{align}
N_{\emptyset\nu}(Q)=\prod_{(i,j)\in\nu}\left(1-Q\,q^{-\nu_{i}+j-1}t^{i} \right).
\end{align}
Upon substituting $Q=v^{2} t^{-n}$ after the identification between the Coulomb branch parameters and masses, the factors in the product will be of the form $(1-q^{-\nu_{i}+j}t^{i-(n+1)})$. For each row $i$ in the Young diagram, the very last box in that row is labelled by $(i,\nu_{i})$ rendering the exponent of $q$ to vanish. On the other hand, if the Young diagram has $(n+1)^{st}$ rows, for that last box the exponent of $t$ vanishes too, hence, the product vanishes. Therefore, the length of that Young diagram is limited to $n$; $\ell(\nu)\leq n$. However, there does not exist any bound on the number of boxes along each of the rows. 
\par{Notice that the contributions from the hypermultiplets are either of the form $N_{\nu\emptyset}(ve/f)$ or $N_{\emptyset\nu}(vf/e)$ and they are in the numerator. On the other hand, the vector multiplet contributions consist of factors $N_{\nu_{a}\nu_{b}}$ in the denominator. It was a crucial point in the derivation of the triality to notice that this type of factors in the denominator can be rewritten as \cite{Aganagic:2013tta},}
\begin{align}\label{degeneration}\nonumber
N_{\nu_{a}\nu_{b}}(Q)&=\prod_{i=1}^{n_{a}}\prod_{j=1}^{n_{b}}\frac{\varphi(Q\,q^{\nu_{a,i}-\nu_{b,j}}t^{j-i+1})}{\varphi(Q\,q^{\nu_{a,i}-\nu_{b,j}}t^{j-i})}\frac{\varphi(Q\,t^{j-i})}{\varphi(Q\,t^{j-i+1})}\\
&\times \left [ N_{\nu_{a}\es}(t^{n_{b}}Q)N_{\es\nu_{b}}(t^{-n_{a}}Q)\right ].
\end{align}
The instanton partition function dramatically simplifies and the remaining factors will be of the following form; 
\begin{align}
N_{\nu\es}(Q)&=\prod_{i=1}^{n}\frac{\varphi(Q\,t^{1-i})}{\varphi(Q\,q^{\nu_{i}}t^{1-i})},\\
N_{\es\nu}(Q)&=\prod_{i=1}^{n}\frac{\varphi(Q\,q^{-\nu_{i}}t^{i})}{\varphi(Q\,t^{i})}.
\end{align} 
These factors written in terms of the quantum dilogarithm nicely combine into the potential of the Dotsenko-Fateev integrals. The remaining factors on the first line in Eq.(\ref{degeneration}), form the Vandermonde determinants. The sum over the Young diagrams can be easily understood as the summation over the residue of the poles encircled by the contours chosen in the integrals,
\begin{align}
\frac{1}{\prod_{a=1}^{M}n_{a}!}\oint_{\vec{\gamma}}d^{n}y\qquad\rightarrow\qquad\sum_{\vec{\nu}},
\end{align}
establishing the direct connection between the conformal blocks and the instanton partition function. We identify the insertion point of the vertex operators $z_{a}=f^{+}_{a}$ and define their momenta by $f^{-}_{a}=q^{\alpha_{a}}f^{+}_{a}v^{-2}$.

\par{In the remainder of the paper, we will argue that this correspondence can be extended to an equivalence between an elliptic deformation of the Dotsenko-Fateev integrals and the 6d instanton partition functions. The elliptic deformation of the conformal blocks is in a similar spirit of the $q$-deformation of the Liouville theory. We subject the vertex operator and the screening charges to an elliptic deformation introduced in \cite{saito}.}

\section{Elliptic Deformation and Its Geometric Interpretation}
\par{In this section, we review the motivation behind the elliptic deformation of \cite{saito} and spell out some details relevant for our approach. Along the way, we recognize a geometric interpretation of this deformation; it corresponds to partially compactifying the web diagram of the toric Calabi-Yau threefold. This compactification of the web lifts the theory from five dimensions (M-theory) to six dimensions (F-theory) and therefore the corresponding Calabi-Yau threefold develops an elliptic fibration. Such compactification of web diagrams was introduced in \cite{Aganagic:2003db} and further studied in \cite{Hollowood:2003cv} to study five dimensional and six dimensional lifts of certain four dimensional theories. In that context, the compactification of the web was associated with the introduction of an new K\"{a}hler class dual to an elliptic curve. Our observation suggests that this appearance of elliptic curve class in the Calabi-Yau threefold is captured by this algebraic deformation at the level of the topological string partition function.}

\subsection{Motivation for the Elliptic Deformation}

\par{The connection between Virasoro/$W_{N}$ algebras and symmetric functions has been known for a while; the Jack symmetric functions are the singular vectors of these algebras. Both sides of this correspondence allows a $q$-deformation; Virasoro/$W_{N}$ algebras are replaced by their $q$-analog as reviewed in the previous section. The $q$-deformation of the Jack symmetric functions are the Macdonald symmetric functions. The Macdonald symmetric functions are defined as eigenfunctions of a difference operator, $H_{N}(q,t)$, henceforth referred to as the Macdonald operator, and defined by}
\begin{align}
H_{N}(q,t)\equiv\sum_{i=1}^{N}\prod_{j\neq i}\frac{tx_{i}-x_{j}}{x_{i}-x_{j}}T_{q,x_{i}},\,\,\mbox{with}\,\,\left (T_{q,x_{i}}f(x_{1},\mathellipsis,x_{N})\equiv f(x_{1},\mathellipsis,qx_{i},\mathellipsis,x_{N})\right).
\end{align}
Macdonald has defined a $(q,t)$-deformed version of the inner product between the power sum symmetric functions (see below) such that it reduced to the usual one when $q=t$, and there is the associate kernel function
\begin{align}
\Pi(x,y;q,t)\equiv \prod_{i,j}\frac{\varphi(tx_{i}y_{j})}{\varphi(x_{i}y_{j})}.
\end{align} 
Based on the kernel function there is a free field representation for the Macdonald operator. This realization is closely related to the Ding-Iohara algebra which whose representations are related to the AGTW conjecture.  
\par{The elliptic version\footnote{In the literature regarding the free field representation of the Macdonald operator $p$ is commonly used as the elliptic parameter. However, in the language of topological string theory $Q_{\tau}$ is more frequently used; we will use both notations interchangeably depending on the context, and hope that it will not create a confusion.} of the Macdonald operator is constructed in \cite{Ruijsenaars}}: 
\begin{align}
H_{N}(q,t,p)\equiv\sum_{i=1}^{N}\prod_{j\neq i}\frac{\Theta_{p}(tx_{i}/x_{j})}{\Theta_{p}(x_{i}/x_{j})}T_{q,x_{i}}.
\end{align}
\par{Later the kernel function for the elliptic Macdonald operator is constructed in \cite{Komori} using the natural generalization of the quantum dilogarithm}
\begin{align}
\Pi(x,y;q,t,p)\equiv\prod_{i,j}\frac{\Gamma(x_{i}y_{j})}{\Gamma(tx_{i}y_{j})},
\end{align}
which reduce to the $qt$-deformed kernel in the limit of $p\rightarrow 0$. In \cite{saito}, an elliptic deformation for the free field representation is presented. Let us review some of the important points of this construction. The vertex operators has generically the following bosonic mode expansion,
\begin{align}
X(z)=\exp\left(\sum_{n>0}X_{-n}^{-} \,a_{-n}\,z^{n}\right)\exp\left(\sum_{n>0}X^{+}_{n}\,a_{n}\,z^{-n} \right).
\end{align}
where $a_{\pm n}$'s are the bosonic generators satisfying the $qt$-deformed version of the usual Heisenberg algebra, 
\begin{align}\label{heisenberg}
[a_{n},a_{m}]=n\frac{1-q^{|n|}}{1-t^{|n|}}\delta_{m+n,0}.
\end{align}

The deformation follows in two steps: first, we deform the Heisenberg algebra for $a_{\pm n}$'s by introducing the elliptic parameter $p$. The deformation alone is not sufficient and need to ``double'' the algebra by launching a new set of generators, denoted by $b_{\pm n}$'s. They satisfy a slightly different Heisenberg algebra. These two sets of bosonic generators are assumed to commute with each other:
\begin{align}\nonumber
&[a_{n},a_{m}]=n(1-p^{|n|})\frac{1-q^{|n|}}{1-t^{|n|}}\delta_{m+n,0},\qquad [b_{n},b_{m}]=n\frac{1-p^{|n|}}{(qt^{-1}p)^{|n|}}\frac{1-q^{|n|}}{1-t^{|n|}}\delta_{m+n,0},\\
&[a_{n},b_{m}]=0.
\end{align}
The second step is constructing the elliptic deformation using the above bosonic generators; $X(z)\equiv X_{b}(p;z)X_{a}(p;z)$ with
\begin{align}
&X_{b}(p;z)\equiv\exp\left(-\sum_{n>0}\frac{p^{n}}{1-p^{n}}X_{n}^{-}\,b_{-n}\, z^{-n}\right)\exp\left(-\sum_{n>0}\frac{p^{n}}{1-p^{n}}X_{-n}^{+}\,b_{n}\,z^n \right),\\
&X_{a}(p;z)\equiv\exp\left(\sum_{n>0}\frac{1}{1-p^{n}}X_{-n}^{-}\,a_{-n}\,z^{n} \right)\exp\left( \sum_{n>0}\frac{1}{1-p^n}X_{n}^{+}\,a_{n}\,z^{-n}\right).
\end{align}

\subsection{Geometric Interpretation of the Elliptic Deformation}

\par{The elliptic deformation that we just discussed is an algebraic construction and we would like to introduce a new geometric interpretation in terms of toric Calabi-Yau threefolds. }

The toric Calabi-Yau threefolds that we are interested in are dual to web diagrams involving $(p,q)$ 5-branes \cite{Leung:1997tw}. These web diagrams in some cases have symmetries and in such cases it is possible to compactify the $\mathbb{R}^2$ in which these webs live to $\mathbb{R}\times S^1$ or $T^2$. Since this $\mathbb{R}^2$ in which the webs live is part of the string theory spacetime therefore such a compactification of the web corresponds to compactification of spacetime in string theory. In case the toric Calabi-Yau threefold gives rise to a certain gauge theory in 4d the compactification of the $\mathbb{R}^2$ to $\mathbb{R}\times S^1$ corresponds to considering M-theory compactified on a toric Calabi-Yau threefold, dual to the compactified web, and hence the corresponding theory is 5d on $\mathbb{R}^{4}\times S^1$. Similarly compactification of the $\mathbb{R}^{2}$ to $T^2$ corresponds to considering F-theory on the toric Calabi-Yau threefold and gives rise to 6d theory on $\mathbb{R}^{4}\times T^2$. The toric Calabi-Yau threefold, dual to the compactified web on $T^2$, in this case must have an elliptic fibration structure. Due to this restriction very few webs exist which can be compactified such that they live on $T^2$. The dual Calabi-Yau threefolds in each of these cases have new elliptic curves whose area is given by the size of the compactification circle.

The web diagrams are dual to the Newton polygon of the Calabi-Yau threefold which encodes the open charts which can be used to build the toric Calabi-Yau threefold. The compactification of the web puts the Newton polygon on a $T^2$ as well. 

\begin{figure}[ht]
  \centering
  \includegraphics[width=1.5in]{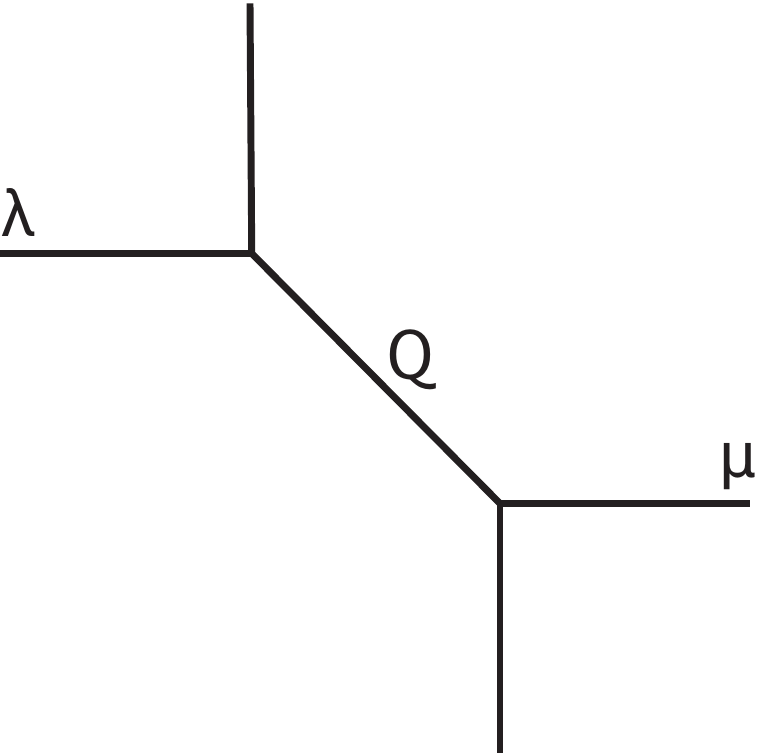}\\
  \caption{The resolved conifold with two infinite stack of branes providing boundary conditions, labelled by Young diagrams $\lambda$ and $\mu$, for open topological string amplitudes.}\label{conifold}
\end{figure}
\par
\begin{figure}[h]
  \centering
  \includegraphics[width=2in]{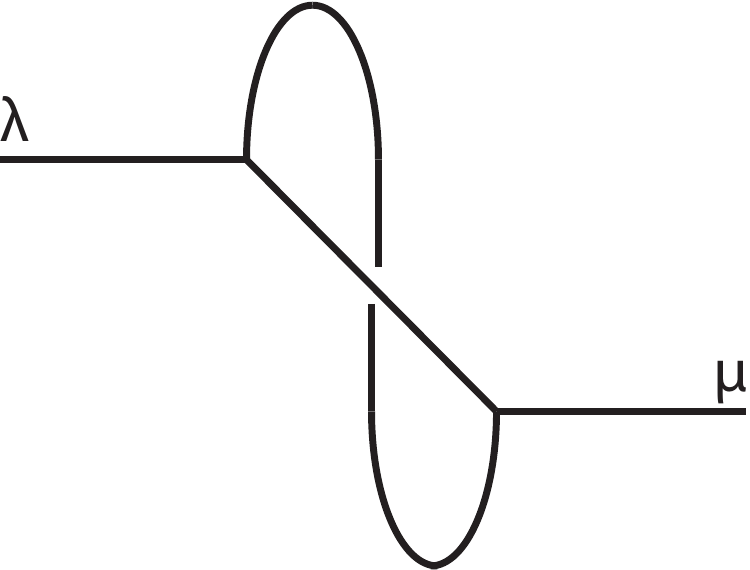}\\
  \caption{The partially compactified web giving rise to 5d gauge theory.}\label{domainwall}
\end{figure}

{The topological vertex formalism \cite{Aganagic:2003db,Iqbal:2007ii} provides an natural way of calculating the topological string amplitudes of the Calabi-Yau threefold dual to the compactified web. In the topological vertex formalism the trivial partitions are associated with the external legs, however, there is a sense of compactification of the geometry by labeling the vertical external legs with the same non-trivial Young diagrams and summing over them. Pictorially, we can think of each external leg as a disc and the Young diagrams will label the boundary conditions for the open maps from the worldsheet. Summing over all possible boundary conditions is in fact gluing two discs along their boundaries, giving rise to a $\mathbb{P}^{1}$. Taking the base $\mathbb{P}^{1}$ of the resolved conifold into account, we end up with two $\mathbb{P}^{1}$'s connected each other at their poles; i.e., a pinched torus. The new web diagram will be non-planar and is depicted in \figref{domainwall}.}
\par{We will establish this conclusion by constructing coherent states labeled by Young diagrams. They will be in the form of vertex operators which we subject to the elliptic deformation in \cite{saito}. The scalar product of two such vectors can be easily computed using the associated $q,t$-deformed Heisenberg algebra. Using the powerful identities in \cite{Awata:2008ed}, we show that the scalar products are precisely the Nekrasov functions, $N_{\lambda\mu}(Q)$. }



\par{We define our coherent states as }
\begin{align}
|\lambda;w\rangle&\equiv\exp\left(\sum_{n=1}^{\infty}\frac{1}{n}\frac{t^{n/2}-t^{-n/2}}{q^{n/2}-q^{-n/2}}w^{n}\varepsilon^{-}_{\lambda}(p_{n})a_{-n}\right)|\emptyset\rangle,\\
\langle w;\lambda |&\equiv\langle\emptyset |\exp\left(\sum_{n=1}^{\infty} \frac{1}{n}\frac{t^{n/2}-t^{-n/2}}{q^{n/2}-q^{-n/2}}w^{-n}\varepsilon_{\lambda}^{+}(p_{n})a_{n}\right).
\end{align}
A number of remarks are in order. The parameters $q$ and $t$ are taken to be pure phase as well as the variable $w$. $a_{\pm n}$ with $n\in{\mathbb Z}\backslash \{0\}$ are creation and annihilation operators and satisfy the $qt$-deformed Heisenberg algebra Eq.(\ref{heisenberg}).
The vector $|\emptyset\rangle$ and its dual $\langle\emptyset |$ represent the vacua and are annihilated by the positive and negative Fourier modes, respectively. We will label the states created by the negative modes as $|a_{\lambda}\rangle=a_{-\lambda_{1}}a_{-\lambda_{2}}\mathellipsis |\emptyset\rangle$, and similarly the dual vectors. We will exploit the isomorphism between the Fock space and the space of symmetric function by identifying $|a_{\lambda}\rangle\sim p_{\lambda}$. Macdonald has defined $(q,t)$-deformation of the inner product on the power sum symmetric functions as 
\begin{align}
\langle p_{\lambda},p_{\mu}\rangle_{q,t}=\delta_{\lambda,\mu}z_{\lambda}\prod_{i=1}^{\ell(\lambda)}\frac{1-q^{\lambda_{i}}}{1-t^{\lambda_{i}}},
\end{align}
where $z_{\lambda}=\prod_{i\geq1}i^{m_{i}}m_{i}!$ with $m_{i}$ being the number of parts of $\lambda$ equal to $i$. It is easy to see that the hermitian conjugate $(a_{-n})^{\dagger}=a_{n}$ with respect to the inner product. 
$\varepsilon_{\lambda}^{\pm}$ is an algebra homomorphism defined on the ring of symmetric functions and its action on the power sums is given by
\begin{align}\nonumber
\varepsilon_{\lambda}^{\pm}(p_{n})&=p_{n}(q^{\pm\lambda}t^{\pm\rho})\\ 
&=\sum_{i=1}^{\infty}q^{\pm\lambda_{i}n}t^{\mp(i-1/2)n}.
\end{align} 
\par{Before we ellipticize the vectors let us show that the transition between two states is proportional to the Nekrasov function. We can commute the two exponentials picking up a factor:}
\begin{align}\nonumber
&\langle z;\lambda |\mu;w\rangle=\\\nonumber
&=\langle\emptyset |\exp\left(\sum_{n=1}^{\infty} \frac{1}{n}\frac{t^{n/2}-t^{-n/2}}{q^{n/2}-q^{-n/2}}z^{-n}\varepsilon_{\lambda}^{+}(p_{n})a_{n}\right)\exp\left(\sum_{n=1}^{\infty}\frac{1}{n}\frac{t^{n/2}-t^{-n/2}}{q^{n/2}-q^{-n/2}}w^{n}\varepsilon^{-}_{\mu}(p_{n})a_{-n}\right)|\emptyset\rangle\\
&=\exp\left(-\sum_{n=1}^{\infty}\frac{1}{n}\left(\frac{w}{z}\right)^{n}v^{n} \frac{t^{-n/2}-t^{n/2}}{q^{n/2}-q^{-n/2}}p_{n}(q^{\lambda}t^{\rho})p_{n}(q^{-\mu}t^{-\rho})\right)
\end{align}
with $v=(q/t)^{1/2}$ and $\rho_{i}={-i+1/2}$ is the Weyl vector. Using the proposition and its corollary in Appendix A.2 of \cite{Awata:2008ed} we can write 
\begin{align}\label{trans}
\langle z;\lambda |\mu;w\rangle&=N_{\lambda\mu}(v^{2} w/z)\prod_{i,j=1}^{\infty}(1-v^{2}w/z\, q^{-j}t^{-i+1})\\\nonumber
&=N_{\lambda\mu}(v^{2} w/z)\langle z;\emptyset |\emptyset;w\rangle.
\end{align}
where the following representation of the Nekrasov function is the more familiar one,
\begin{align}
N_{\lambda\mu}(Q)=\prod_{(i,j)\in\lambda}\left(1-Q\,q^{\lambda_{i}-j}t^{\mu_{j}^{t}-i+1} \right)\prod_{(i,j)\in\mu}\left(1-Q\,q^{-\mu_{i}+j-1}t^{-\lambda_{j}^{t}+i} \right).
\end{align}
\par
\begin{figure}[h]
  \centering
  \includegraphics[width=3in]{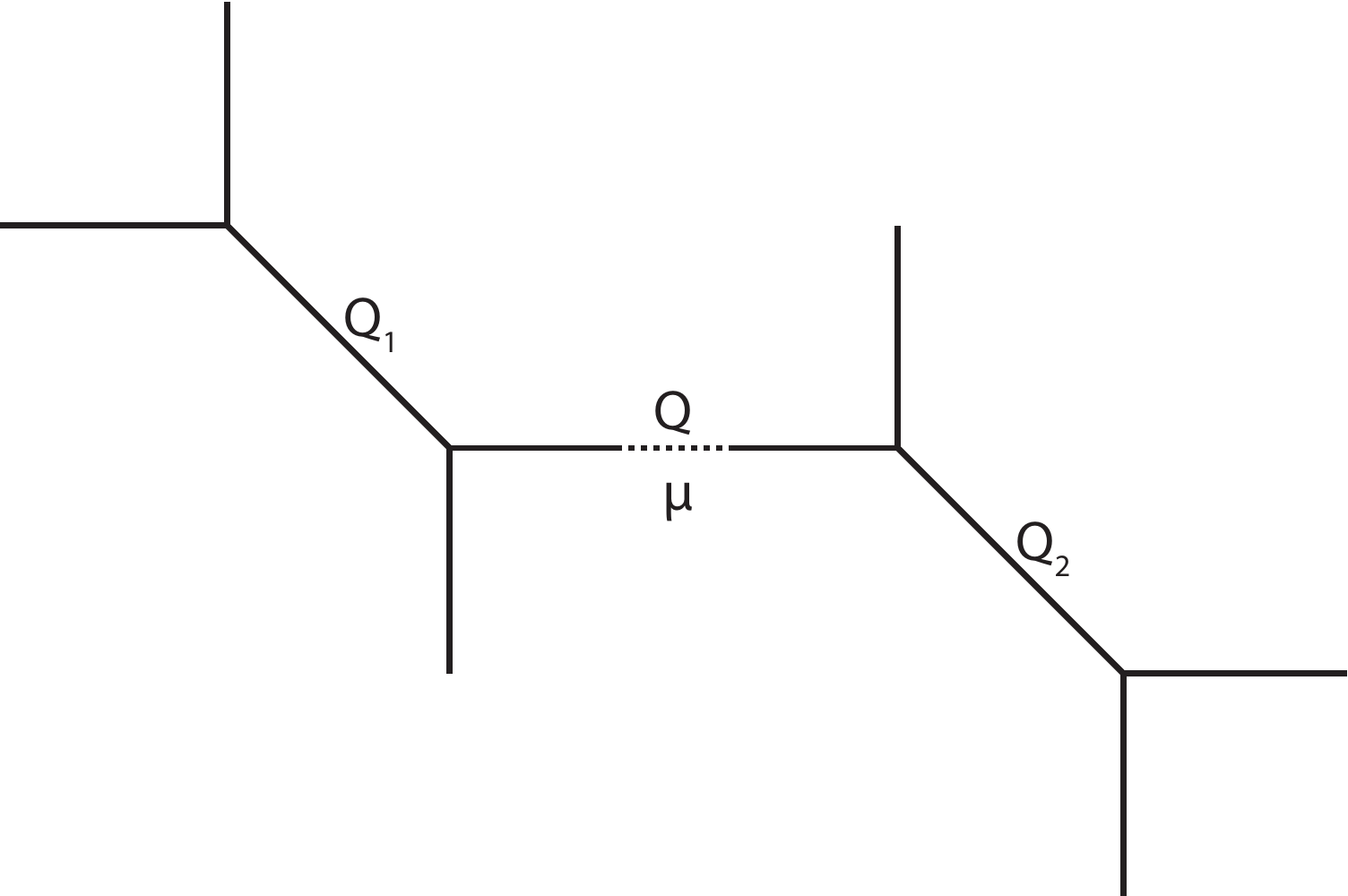}\\
  \caption{The web dual to the resolved conifold blown up at two points which gives $U(1)$ theory with two hyper multiplets.}\label{conifoldblownup}
\end{figure}

{As we mentioned before, the Nekrasov function is not the same as the normalized open topological string amplitude on the resolved conifold. We can compute it using the refined topological vertex to be}
\begin{align}\nonumber
{\cal Z}_{\lambda\mu}&=\sum_{\nu}(-Q)^{|\nu|}C_{\emptyset\nu\lambda^{t}}(q^{-1},t^{-1})C_{\emptyset\nu^{t}\mu}(t^{-1},q^{-1})\\\nonumber
&={\cal Z}_{\emptyset\emptyset}\,q^{-\Arrowvert\mu\Arrowvert^{2}/2}{\widetilde Z}_{\mu}(t^{-1},q^{-1})t^{-\Arrowvert\lambda^{t}\Arrowvert^{2}/2}{\widetilde Z}_{\lambda^{t}}(q^{-1},t^{-1})N_{\lambda\mu}(Q)\\
&={\cal Z}_{\emptyset\emptyset}\,(-1)^{|\lambda|+|\mu|}v^{|\lambda|-|\mu|}P_{\mu}(t^{-\rho};q,t)P_{\lambda^{t}}(q^{-\rho};t,q)N_{\lambda\mu}(Q)
\end{align}
where ${\cal Z}_{\emptyset\emptyset}$ is the closed amplitude by setting the representations along the external legs to trivial representations $\emptyset$ and the factors ${\widetilde Z}_{\mu}(t,q)$ are related to Macdonald functions with special arguments $P_{\mu}(t^{-\rho};q,t)$,
\begin{align}\nonumber
{\widetilde Z}_{\mu}(t,q)&=\prod_{(i,j)\in\mu}\frac{1}{1-q^{\mu_{i}-j} t^{\mu^{t}_{j}-i+1}}\\
&=t^{-\Arrowvert\mu^{t}\Arrowvert^{2}/2}P_{\mu}(t^{-\rho};q,t).
\end{align}
How about the remaining factors? Do they have any connection to the coherent states we constructed? To answer these question, let us compute the partition function of the resolved conifold blown up at two points, whose toric diagram is sketched in \figref{conifoldblownup}. 

\par{The (normalized) topological string partition function takes the following form in terms of the Nekrasov functions: }
\begin{align}\nonumber
Z&=\sum_{\mu}\left(Qv \right)^{|\mu|}\frac{N_{\emptyset\mu}(Q_{1}v)N_{\mu\emptyset}(Q_{2}v)}{N_{\mu\mu}(v^2)}\\
&=\sum_{\mu}\left(Qv\right)^{|\mu|}\frac{\langle\emptyset;z_{1}|\mu;w\rangle\langle\mu;w|\emptyset;z_{2}\rangle}{\langle w;\mu|\mu;w\rangle},
\end{align}
where $Q_{1}=vw/z_{1}$ and $Q_{2}=vz_{2}/w$ are exponentials of the masses of hypermultiplets\footnote{The K\"{a}hler parameters seem to be identified slightly different; it is related to the particular geometry chosen, but related by a flop transition.}. We also have ignored the closed string partition function used in the normalization of the inner product Eq.(\ref{norm}). The pre-factors from two pieces combine into to the norm of our coherent states:

\begin{align}\label{norm}
\frac{\langle w;\mu |\mu;w\rangle}{\langle w;\emptyset |\emptyset;w\rangle}=N_{\mu\mu}(v^{2}).
\end{align}

\subsection{A Closer Look to the Coherent States}
Before we move on to computing the elliptic deformation of our coherent states, let us point out few interesting features of them. According to the Proposition 2.14 in \cite{Awata:2011dc}, this inner product is proportional to the matrix elements of the following vertex operator in the integral Macdonald basis,
\begin{align}
\langle J_{\lambda}|\Phi(w)|J_{\mu}\rangle=N_{\lambda\mu}(qv/tu)w^{|\lambda|-|\mu|}(tu/q)^{|\lambda|}(-v/q)^{-|\mu|}t^{n(\lambda)}q^{n(\mu^{t})}.
\end{align}
where the vertex operator is defined as
\begin{align}
\Phi(w)=\exp\left(-\sum_{n=1}^{\infty}\frac{1}{n}\frac{v^{n}-(t/q)^{n}u^{n}}{1-q^{n}}a_{-n}w^{n}\right)\exp\left(\sum_{n=1}^{\infty}\frac{1}{n}\frac{v^{-n}-u^{-n}}{1-q^{-n}}a_{n}w^{-n} \right).
\end{align}

The elliptic deformation is suitable for operator given in terms of Fourier modes, although the vertex operator $\Phi(w)$ is given in this form, computing the deformation of the integral Macdonald functions is a very interesting but a challenging task. We will postpone this discussion for another place, but assume the elliptic deformation of our vectors and their inner product capture the relevant deformation for these matrix elements as well. 

\par{We can generalize our vectors to capture not only the normalized partition function of the resolved conifold but the so-called strip geometry. Instead of labeling each state with one Young diagram, corresponding to the external leg on the toric diagram, we have $r$ diagrams. We can use a collective notation $\vec{\lambda}$ to label the $r$-tuple of Young diagrams, and $\vec{z}$ to label the associated parameters:   }
\begin{align}
|\vec{\lambda};\vec{w}\rangle&\equiv\exp\left(\sum_{n=1}^{\infty}\frac{1}{n}\frac{t^{n/2}-t^{-n/2}}{q^{n/2}-q^{-n/2}}\sum_{i=1}^{r}w_{i}^{n}\varepsilon^{-}_{\lambda_{i}}(p_{n})a_{-n}\right)|\emptyset\rangle.
\end{align}
It is not hard to show that the inner product between two such states reproduce the correct topological string partition function
\begin{align}
\langle\vec{z};\vec{\lambda}|\vec{\mu};\vec{w}\rangle=\prod_{i,j=1}^{r}N_{\lambda_{i}\mu_{j}}(vw_{j}/z_{i}).
\end{align}
\begin{figure}[h]
  \centering
  \includegraphics[width=1.5in]{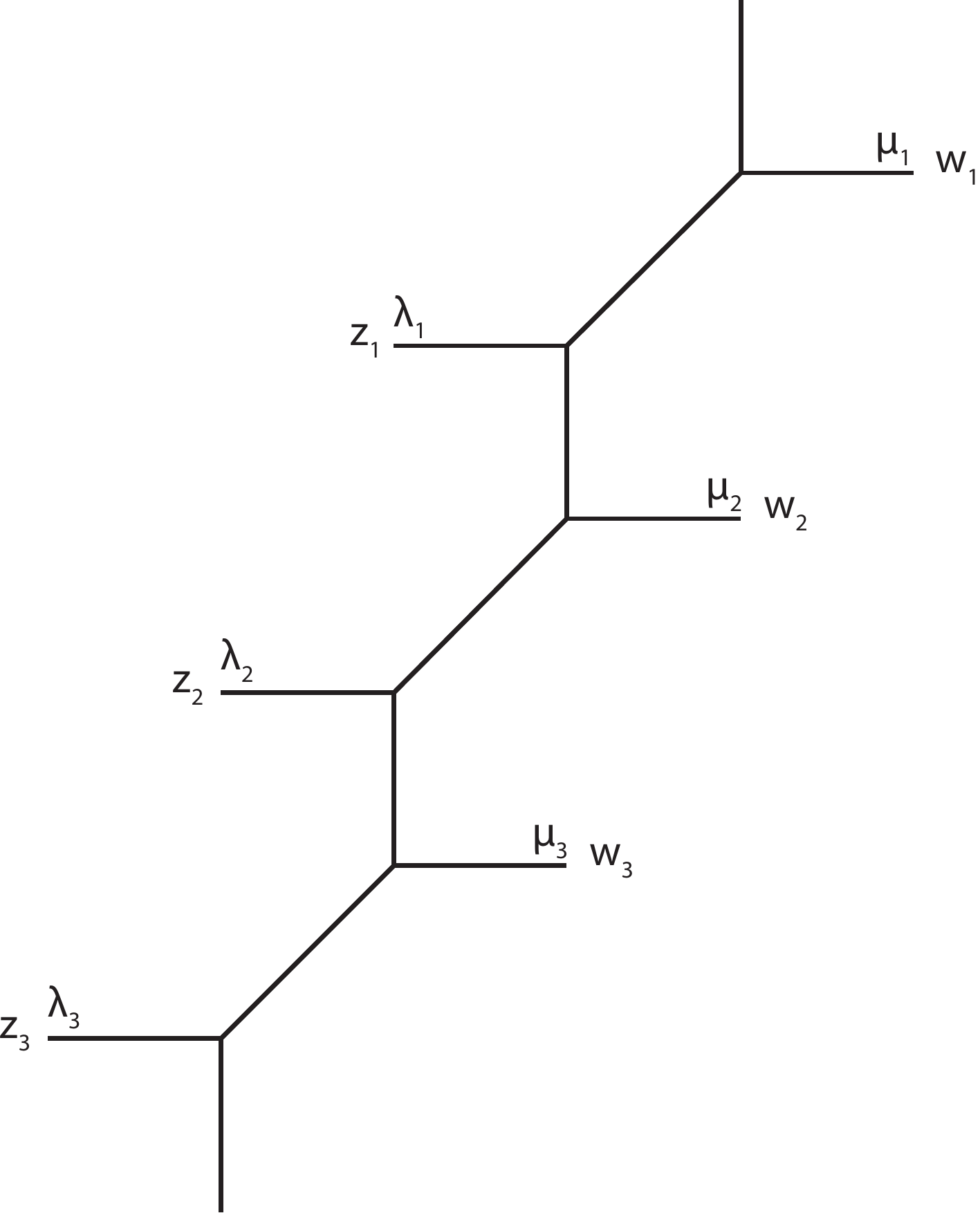}\\
  \caption{The strip geometry for $r=3$.}\label{strip}
\end{figure}
One can easily see that the partition function is getting contributions from the holomorphic maps in the geometry.

\par{We have defined the vectors and computed their inner products. We can write them in the so-called Macdonald basis and compute the inner product using this basis as well. The kernel function for the Macdonald polynomials is given as}
\begin{align}
\exp\left(\frac{1}{n}\frac{1-t^{n}}{1-q^{n}}p_{n}(x)p_{n}(y) \right)=\sum_{\lambda}P_{\lambda}(x;q,t)Q_{\lambda}(y;q,t),
\end{align}
where $P_{\lambda}(x;q,t)$ is the Macdonald function and $Q_{\lambda}(y;q,t)$ is its dual with respect to the $qt$-deformation of the inner product
\begin{align}
\langle P_{\lambda},Q_{\mu}\rangle_{q,t}=\delta_{\lambda,\mu}.
\end{align}
For later convenience let us also mention another important relation between them,
\begin{align}
Q_{\lambda}=\frac{1}{\langle P_{\lambda},P_{\lambda}\rangle_{q,t}}P_{\lambda}\equiv b_{\lambda} P_{\lambda},
\end{align}
where $b_{\lambda}=b_{\lambda}(q,t)$ is given as a product over the boxes of the Young diagram $\lambda$
\begin{align}
b_{\lambda}(q,t)=\prod_{s\in\lambda}\frac{1-q^{a(s)}t^{l(s)+1}}{1-q^{a(s)+1}t^{l(s)}}.
\end{align}
We can act on both sides with the operator $\varepsilon^{-}_{\lambda}$ to the $y$-variables and obtain after using the isomorphism between the symmetric functions and the Fock space
\begin{align} \nonumber
|\lambda;w\rangle&=\exp\left(\sum_{n=1}^{\infty}\frac{1}{n}\frac{t^{n/2}-t^{-n/2}}{q^{n/2}-q^{-n/2}}w^{n}\varepsilon^{-}_{\lambda}(p_{n})a_{-n}\right)|\emptyset\rangle\\
&=\sum_{\sigma}(vw)^{|\sigma|}Q_{\sigma}(q^{-\lambda}t^{-\rho};q,t)|P_{\sigma}\rangle,
\end{align}
where $|P_{\sigma}\rangle$ denote the vectors in the Macdonald basis. We can take the Hermitian conjugate of both sides to get
\begin{align}
\langle w;\lambda|=\sum_{\sigma}w^{-|\sigma|}v^{|\sigma|}Q_{\sigma}(q^{\lambda}t^{\rho};q,t)\,\langle P_{\sigma}|,
\end{align}
where we have used $Q_{\sigma}(x;q^{-1},t^{-1})=(qt^{-1})^{|\sigma|}Q_{\sigma}(x;q,t)$. Let us compute the inner product between two states in the Macdonald basis
\begin{align}\nonumber
\langle z;\lambda|\mu;w\rangle&=\sum_{\sigma}\frac{1}{b_{\sigma}(q,t)}(w/z)^{|\sigma|}v^{2|\sigma|} Q_{\sigma}(q^{\lambda}t^{\rho};q,t)Q_{\sigma}(q^{-\mu}t^{-\rho};q,t)\\\nonumber
&=\Pi(v^{2}w/z\,q^{\lambda}t^{\rho},q^{-\mu}t^{-\rho};q,t)\\
&=\Pi(v^{2}w/z\,t^{\rho},t^{-\rho};q,t)N_{\lambda\mu}(v^{2}w/z),
\end{align}
which precisely agrees with Eq.(\ref{trans}) using Eq.(A.24) of \cite{Awata:2008ed}.

\subsection{Elliptic Deformation}

Using the prescription in \cite{saito}, we can compute the elliptic deformation of our vectors, denoted with boldfaces to distinguish them from non-elliptic ones, and their inner products
\begin{align}
\pmb{\langle\lambda;z|\mu;w\rangle}=\prod_{i,j=1}^{\infty}\frac{\Gamma\left(v^{2} w/z\, q^{\lambda_{i}-\mu_{j}}t^{j-i}\right)}{\Gamma\left(v^2 w/z\, q^{\lambda_{i}-\mu_{j}}t^{j-i+1}\right)},
\end{align}
where $\Gamma(x)$ is the elliptic gamma function defined as 
\begin{align}
\Gamma(x)=\prod_{m,n=0}^{\infty}\frac{1-x^{-1}q^{m+1}p^{n+1}}{1-xq^{m}p^{n}}.
\end{align}
We use bold face for the elliptic deformation of the coherent states to distinguish them from the regular ones. At first sight, the elliptic deformation might not look anything like the partition function of the building blocks $D_{\lambda\mu}$ for M-strings, and likewise the Nekrasov function $N_{\lambda\mu}(Q)$. For completeness, let us remind the building blocks $D_{\lambda\mu}$
\begin{align}\nonumber
&D_{\lambda\mu}=t^{-\frac{\Arrowvert\mu^{t}\Arrowvert^2}{2}}\,q^{-\frac{\Arrowvert\lambda\Arrowvert^2}{2}}Q_{m}^{-\frac{|\lambda|+|\mu|}{2}}\\ \nn &\times\prod_{k=1}^{\infty}\prod_{(i,j)\in\lambda}\frac{(1-Q_{\tau}^{k}Q_{m}^{-1}\,q^{-\lambda_{i}+j-\frac{1}{2}}\,t^{-\mu_{j}^{t}+i-\frac{1}{2}})
(1-Q_{\tau}^{k-1}Q_{m}\,q^{\lambda_{i}-j+\frac{1}{2}}\,t^{\mu_{j}^{t}-i+\frac{1}{2}})}{(1-Q_{\tau}^{k}\,q^{\lambda_{i}-j}\, t^{\lambda_{j}^{t}-i+1})(1-Q_{\tau}^{k-1}\,q^{-\lambda_{i}+j-1}\,t^{-\lambda_{j}^{t}+i})}\\
&\times\prod_{(i,j)\in\mu}\frac{(1-Q_{\tau}^{k}Q_{m}^{-1}\,q^{\mu_{i}-j+\frac{1}{2}}t^{\lambda_{j}^{t}-i+\frac{1}{2}})(1-Q_{\tau}^{k-1}Q_{m} \,q^{-\mu_{i}+j-\frac{1}{2}}t^{-\lambda_{j}^{t}+i-\frac{1}{2}})}{(1-Q_{\tau}^{k}\,q^{\mu_{i}-j+1}t^{\mu_{j}^{t}-i})(1-Q_{\tau}^{k-1}\,q^{-\mu_{i}+j}t^{-\mu_{j}^{t}+i-1})}. \label{eq:domainwall}
\end{align}

However, the following identity from \cite{Awata:2008ed} allows us to prove that the numerator of $D_{\lambda\mu}$ is given by the elliptic deformation of the transition between our coherent states:
\begin{align}\nonumber
\sum_{(i,j)\in\lambda}q^{\lambda_{i}-j+1/2}&\,t^{\mu_{j}^{t}-i+1/2}+\sum_{(i,j)\in\mu}q^{-\mu_{i}+j-1/2}t^{-\lambda_{j}^{t}+i-1/2}\\ 
&=\frac{t^{-1/2}-t^{1/2}}{q^{1/2}-q^{-1/2}}\left(\sum_{i=1}^{\infty}q^{\lambda_{i}}t^{-i+1/2}\sum_{j=1}^{\infty}q^{-\mu_{j}}t^{j-1/2}-\sum_{i=1}^{\infty}t^{-i+1/2}\sum_{j=1}^{\infty}t^{j-1/2} \right).
\end{align}
We take $q\rightarrow q^{k}$, $t\rightarrow t^{k}$, place the above expressions into $\{\mathellipsis\}$ and perform the sums:
\begin{align}
-\sum_{k=1}^{\infty}\frac{Q^{k}}{k}\frac{p^k}{1-p^{k}}\left\{\mathellipsis\right\}.
\end{align}
The denominator is again related to the norm of the states. For example, the M-string partition function in the presence of two M5-branes is precisely the partial compactification of the resolved conifold blown up at two points. Its partition function turns out to be expressed in terms of the elliptic version of our coherent states in the same form:
\begin{align}
Z=\sum_{\lambda}\frac{\pmb{\langle\emptyset;z|\lambda;w\rangle}\pmb{\langle\lambda;w|\emptyset;z\rangle}}{\pmb{\langle\lambda;w|\lambda;w\rangle}}Q_{f}^{|\lambda|}
\end{align}
where $Q_{m} \equiv vw/z$. We conclude that \textit{the elliptic deformation corresponds to a partial compactification of the web diagram.} Note that the elliptic deformation implicitly leave the instanton direction untouched, and affects the ``perpendicular'' direction to it. We also ignore some irrelevant factors that do not depend on the representations.

\section{Elliptic Deformation of Conformal Blocks}
\par{In this section we want to apply the elliptic deformation to the correspondence between the $q$-deformed conformal blocks and the instanton partition function, introduced in \cite{Aganagic:2013tta} for $A_{1}$ and subsequently generalized to $A_{r}$ and $DE$-groups in \cite{Aganagic:2014oia, Aganagic:2015cta}.   }
\par{The deformation of the $W_{q,t}$ algebras associated to simple Lie algebras are introduced by \cite{FR}, generalizing earlier works of \cite{Shiraishi1,Shiraishi2}. The Heisenberg algebra generators are coupled to each other using the associated Cartan matrix of the Lie algebra. We will change the Heisenberg algebra in \cite{Aganagic:2014oia} to apply the elliptic deformation that we are using. The screening charges $S_{(i)}(z)$ were defined in \cite{Aganagic:2014oia} as}
\begin{align}
S_{(i)}(z)= \,\,:\exp\left(\sum_{k\neq0}\frac{{\tilde\alpha}_{n}^{(i)}z^{-n}}{q^{n/2}-q^{-n/2}}\right):
\end{align}
where the Fourier modes are defined in terms of generators of $r+1$ commuting copies of Heisenberg algebras:
\begin{align}\nonumber
{\tilde \alpha}_{n}^{(i)}&=\frac{1}{n}(t^{-n/2}-t^{n/2})(v^{-n}h_{n}^{(i)}-h_{n}^{(i+1)}),\qquad (k>0),\\
{\tilde \alpha}_{-n}^{(i)}&=\frac{1}{n}(q^{-n/2}-q^{n/2})(h_{-n}^{(i)}-v^{n}h_{-n}^{(i+1)}),\qquad (k>0).
\end{align}
The Heisenberg algebra generators $h_{n}^{(i)}$ satisfy the following relations
\begin{align}
[h_{n}^{(i)},h_{m}^{(j)}]=n\delta_{n+m,0}\delta^{i,j},\qquad \mbox{for}\,\,i,j=1,\mathellipsis,r+1\,\, \mbox{and}\,\, n,m\in\mathbb{Z}.
\end{align}
Instead of the above Heisenberg generators $h_{n}^{(i)}$, we generalize the Eq.(\ref{heisenberg}) to
\begin{align}
[a_{n}^{(i)},a_{m}^{(j)}]=n\frac{1-q^{|n|}}{1-t^{|n|}}\delta_{n+m,0}\delta^{i,j}.
\end{align}
The transformation between these sets of generators can be explicitly realized also as
\begin{align}
a_{n}^{(i)}=h_{n}^{(i)},\,\,\qquad\mbox{and}\qquad a_{-n}^{(i)}=\frac{1-q^{n}}{1-t^{n}}h_{-n}^{(i)},\qquad n>0.
\end{align}
This change is not only consistent with our notation but allows us to write the Fourier modes of the screening charges more compact. The modes are given by
\begin{align}
\alpha_{n}^{(i)}=\frac{1-t^{|n|}}{|n|}\left(q^{-|n|/2}a_{n}^{(i)}-t^{-|n|/2} a_{n}^{(i+1)}\right),\qquad\mbox{for}\,\, k\neq0.
\end{align} 
It is easy to see that they satisfy the same commutation relations as ${\tilde \alpha}_{n}^{(i)}$
\begin{align}\nonumber
[\alpha_{n}^{(i)},\alpha_{m}^{(j)}]&=[{\tilde\alpha}_{n}^{(i)},{\tilde\alpha}_{m}^{(j)}]\\
&=\frac{1}{n}\delta_{n,m}(t^{-n/2}-t^{n/2})(q^{-n/2}-q^{n/2})\left\{ (v^{n}+v^{-n})\delta^{i,j}-(\delta^{i+1,j}+\delta^{i,j+1})\right\},
\end{align}
where the factor depending on the roots is easily seen to be the $qt$-deformation of the Cartan matrix of $A_{r}$ \cite{FR}. The screen charges with respect to our modes is, \begin{align}
S_{(i)}(z)= \,\,:\exp\left(\sum_{k\neq0}\frac{\alpha_{n}^{(i)}z^{-n}}{q^{n/2}-q^{-n/2}}\right):.
\end{align}
\par{In addition to the screening charges, we need the primary vertex operator. The vertex operator in \cite{Aganagic:2014oia} can be rewritten in terms of new modes as}
\begin{align}
V_{\alpha}(z)=\,\,:\exp\left(\sum_{n\neq0}\sum_{i=1}^{r+1}\frac{1}{n}\frac{1}{1-q^{n}}q^{n\alpha_{i}}a_{n}^{(i)}z^{-k} \right):.
\end{align}
This $q$-deformed vertex operator does not have a well defined conformal limit, $q\rightarrow1$. However, this pathology can be cured by adding appropriate additional terms in the exponent. It turns out these additional terms commute with the screening charges. As long as we are concerned with the Dotsenko-Fateev integral and the two point functions the above form should be sufficient\footnote{We would like to thank N. Haouzi for useful discussions.}.  Having defined the screening charges and the primary vertex operators we can elliptically deform. The two point functions which are relevant to compute the Dotsenko and Fateev integrals become
\begin{align}\nonumber
\langle \pmb{S_{(i)}(y)S_{(i)}(y')}\rangle&= \frac{\Gamma(ty'/y)\Gamma(qy'/y)}{\Gamma(y'/y)\Gamma(qt^{-1}y'/y)}\\ \nonumber
\langle \pmb{S_{(i)}(x)S_{(i+1)}(y)}\rangle&=\frac{\Gamma(uy/x)}{\Gamma(vy/x)},\\
\langle \pmb{V_{\alpha}(z)S_{(i)}(x)}\rangle&=\frac{\Gamma(pv^{-1}q^{1-\alpha_{i+1}}z/x)}{\Gamma(pq^{1-\alpha_{i}}z/x)},
\end{align}
where we followed our notation indicating the elliptic deformation by boldfaces. Although we have computed the two point function for the $A_{r}$, we will specialize only to the rank one algebra.  

\section{6d Instanton Partition Function and Elliptic Conformal Blocks}
\par{In this section we will compute the instanton partition function of 6d gauge theories with 8 supercharges. We will focus on a very particular class of theories that can be engineered upon compactification of the non-compact toric Calabi-Yau threefolds. The refined topological vertex can be used to compute their instanton partition functions. Some details are outlined in the Appendix B of \cite{Haghighat:2013gba}. In general, F-theory on an elliptically fibered Calabi-Yau threefold with a section gives rise to 6d theories. For the 6d theory to be anomaly free a specific matter content is needed. In this paper, the compactified 6d $U(N)$ theory with $N_{f}=2N$ hypermultiplets are considered which arise from a toric elliptically fibered Calabi-Yau threefold. }


\subsection{6d $U(1)$ Theory with $N_{f}=2$}
\par
\begin{figure}[h]
  \centering
  \includegraphics[width=1in]{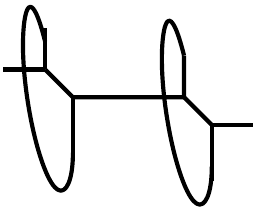}\\
  \caption{The web which gives six dimensional $U(1)$ theory with $N_f=2$.}\label{6dn2}
\end{figure}

{The web diagram which gives this gauge theory is depicted in \figref{6dn2}. The refined topological vertex can be implemented to compute the instanton partition function. In \cite{Haghighat:2013gba} this was calculated and the topological string partition function was expressed in terms of Jacobi $\theta$-functions. However, we write down the partition function in terms of Nekrasov function to facilitate the truncation almost identically to the un-elliptic case,  }

\begin{align}\nonumber
{\cal Z}=\sum_{\nu}(vQ_{b})^{|\nu|}\prod_{k=1}^{\infty}&N_{\nu\es}(\qt^{k-1}\,e/f^{-})N_{\es\nu}(v\qt^{k}\,f^{-}/e)N_{\es\nu}(v\qt^{k-1}\,f^{+}/e)N_{\nu\es}(v\qt^{k}\,e/f^{+})\\
\times&N_{\nu\nu}^{-1}(v^{2}\qt^{k-1})N_{\nu\nu}^{-1}(\qt^{k}).
\end{align}
Although this is the simplest example we will consider, it features the essential ingredients for the more general case; hence, we want to be more explicit. At the point when the Coulomb branch meets the Higgs branch, we impose again the following condition relating the ``Coulomb branch parameter'' to the mass of the fundamental hypermultiplet
\begin{align}
e=\frac{t^{n}}{v}f^{+}.
\end{align}
First of all, note that this identification again puts a restriction to the length of the Young diagrams we are summing over. The restriction is enforced by the Nekrasov function $N_{\es\nu}(v\qt^{k-1}\,f^{+}/e)$ when $k=1$. Using Eq.(\ref{degeneration}), we see that all the contributions coming from the hypermultiplet in the fundamental representation cancel against factors coming from splitting the Nekrasov functions of the vector multiplet, and the partition function reduces to 
\begin{align}\nonumber
{\cal Z}&=\sum_{\nu}(vQ_{b})^{|\nu|}\prod_{k=1}^{\infty}\prod_{i,j=1}^{n}\frac{\varphi(v^{2}\qt^{k-1}\,q^{\nu_{i}-\nu_{j}}t^{j-i})}{\varphi(v^{2}\qt^{k-1}\,q^{\nu_{i}-\nu_{j}}t^{j-i+1})}\frac{\varphi(\qt^{k}\,q^{\nu_{i}-\nu_{j}}t^{t^{j-i}})}{\varphi(\qt^{k}\,q^{\nu_{i}-\nu_{j}}t^{j-i+1})}\frac{\varphi(v^{2}\qt^{k-1}\,t^{j-i+1})}{\varphi(v^{2}\qt^{k-1}\,t^{j-i})}\\
&\times\frac{\varphi(\qt^{k}\,t^{j-i+1})}{\varphi(\qt^{k}\,t^{t^{j-i}})}\frac{N_{\nu\es}(\qt^{k-1}\,t^{n}f^{+}/f^{-})N_{\es\nu}(v^{2}\qt^{k}\,t^{-n}f^{-}/f^{+})}{N_{\nu\es}(v^{2}\qt^{k-1}\,t^{n})N_{\es\nu}(\qt^{k}\,t^{-n})}.
\end{align}
After the analytical continuation performed in \cite{Aganagic:2014oia}, it is easy to show that the quantum dilogarithm functions combine into elliptic $\Gamma$-functions, matching the elliptic Vandermonde determinant defined by the elliptically deformed screening charges. Similarly, the potential can be shown to match the two point function between the vertex operator and screening charge:  
\begin{align}\nonumber
\frac{\left[V(y)\right]_{y=y_{\nu}}}{\left[V(y)\right]_{y=y_{\es}}}=\prod_{i=1}^{n}&\frac{\Gamma(q^{-\nu_{i}}t^{-n+i})\Gamma(v^{2}(f^{-}/f^{+})t^{-n+i})}{\Gamma(v^2(f^{-}/f^{+})\,q^{-\nu_{i}}t^{-n+i})\Gamma(t^{-n+i})},
\end{align} 
with the same identifications as before about the insertions points and momenta. We have also gotten rid of $\qt$ in the argument of gamma function by factoring out an irrelevant factor for the instanton computation. Our result obviously reproduces the $q$-deformed Liouville/5d connection in the limit $\qt\rightarrow 0$; since $\Gamma(x)\mapsto\varphi^{-1}(x)$ in this limit.
\par{We want to emphasize an important point about the contours; one may ask the question if the contours need to be deformed since we have replaced all quantum dilogarithms with elliptic gamma functions. The integrand definitely has more poles! We claim that the contours are \textit{unchanged}, the same way that the contours remained untouched when the $q$-deformation was introduced to the Liouville theory. Our answer is motivated by the exact match of the elliptic conformal blocks to the 6d instanton partition function. In addition, the meaning of  $q$ and $p$ are very different from the point of view of topological string theory and localization computations, although the elliptic gamma functions are symmetric under their exchange. We just lifted the theory one dimension up, but we are still performing the same localization computation on the same $\Omega$-background. There does not seem to be any reason why these additional poles should be incorporated.   }

\subsection{6d $U(2)$ Theory with $N_{f}=4$ }
\par{In this section, we would like to demonstrate that the correspondence holds for 6d $U(2)$ theory with $N_{f}=4$. The web diagrams of the 5d and its compactification which gives the 6d theory are shown in \figref{6dn4}. 

\begin{figure}[h]
  \centering
  \includegraphics[width=1.3in]{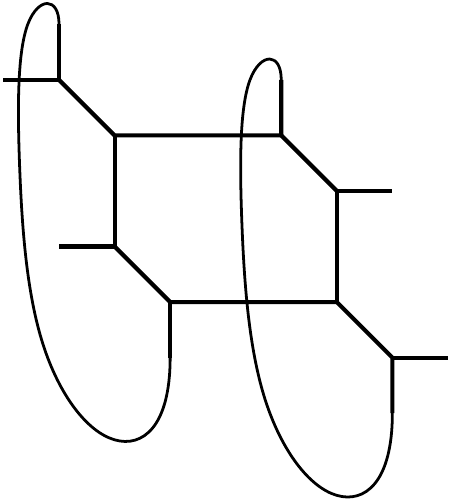}\\
  \caption{The web giving the six dimensional $U(2)$ theory with $N_{f}=4$.}\label{6dn4}
\end{figure}

The topological string partition function can again be expressed in terms of Jacobi theta functions or infinite product of Nekrasov factors and is given by:}
\begin{align}\nonumber
&{\cal Z}=\sum_{\nu_{1},\nu_{2}}(vQ_{b_{1}})^{|\nu_{1}|}(vQ_{b_{2}})^{|\nu_{2}|}\prod_{k=1}^{\infty}N_{\es\N}(v\qt^{k}\, f_{1}^{-}/e_{1})N_{\N\es}(v\qt^{k-1}\,e_{1}/f_{1}^{-})N_{\es\N}(v\qt^{k-1}\,f_{2}^{-}/e_{1})\\ \nonumber
&\times N_{\N\es}(v\qt^{k}\,e_{1}/f_{2}^{-})N_{\es\NN}(v\qt^{k}\,f_{1}^{-}/e_{2})N_{\NN\es}(v\qt^{k-1}\,e_{2}/f_{1}^{-})N_{\NN\es}(v\qt^{k-1}\,e_{2}/f_{2}^{-})N_{\es\NN}(v\qt^{k}\,f_{2}^{-}/e_{2})\\ \nonumber
&\times N_{\N\es}(v\qt^{k}\,e_{1}/f_{2}^{+})N_{\es\N}(v\qt^{k-1}\,f_{2}^{+}/e_{1})N_{\es\N}(v\qt^{k-1}\,f_{1}^{+}/e_{1})N_{\N\es}(v\qt^{k}\,e_{1}/f_{1}^{+})N_{\es\NN}(v\qt^{k-1}\,f_{2}^{+}/e_{2})\\ \nonumber
&\times N_{\NN\es}(v\qt^{k}\,e_{2}/f_{2}^{+})N_{\NN\es}(v\qt^{k-1}\,e_{2}/f_{1}^{+})N_{\es\NN}(v\qt^{k}\,f_{1}^{+}/e_{2})N^{-1}_{\N\N}(v^{2}\qt^{k-1})N^{-1}_{\N\N}(\qt^{k})N^{-1}_{\NN\NN}(v^{2}\qt^{k-1})\\ 
&\times N^{-1}_{\NN\NN}(\qt^{k})N^{-1}_{\N\NN}(v^{2}\qt^{k}\,e_{1}/e_{2})N^{-1}_{\NN\N}(v^{2}\qt^{k-1}\,e_{2}/e_{1})N^{-1}_{\N\NN}(\qt^{k}\,e_{1}/e_{2})N^{-1}_{\NN\N}(\qt^{k-1}\,e_{2}/e_{1})
\end{align}
The Coulomb branch parameters are tuned to be the same as the masses of half of the multiplets up to integer multiple of $\epsilon_{1}$,
\begin{align}
e_{i}=\frac{t^{n_{i}}}{v}f_{i}^{+},\qquad\qquad\mbox{for}\,\, i=1,2.
\end{align}
It is almost identical to the previous case to show that the elliptic Vandermonde is produced after truncation. The remaining factors can be grouped to give rise to the potentials,
\begin{align}\nonumber
\frac{\left[V_{1}(y)\right]_{y=y_{\N}}}{\left[V_{1}(y)\right]_{y=y_{\es}}}=\prod_{i=1}^{n_{1}}&\frac{\Gamma(q^{-\nu_{i,1}}t^{-n_{1}+i})\Gamma((f^{+}_{2}/f_{1}^{+})\,q^{-\nu_{i,1}}t^{-n_{1}+i})}{\Gamma(v^2 (f_{1}^{-}/f_{1}^{+})\,q^{-\nu_{i,1}}t^{-n_{1}+i})\Gamma(v^2 (f_{2}^{-}/f_{1}^{+})\,q^{-\nu_{i,1}}t^{-n_{1}+i})}\\
\times&\frac{\Gamma(v^2 (f_{1}^{-}/f_{1}^{+})\,t^{-n_{1}+i})\Gamma(v^2 (f_{2}^{-}/f_{1}^{+})\,t^{-n_{1}+i})}{\Gamma(t^{-n_{1}+i})\Gamma((f^{+}_{2}/f_{1}^{+})\,t^{-n_{1}+i})},
\end{align} 
and,
\begin{align}\nonumber
\frac{\left[V_{2}(y)\right]_{y=y_{\NN}}}{\left[V_{2}(y)\right]_{y=y_{\es}}}=\prod_{i=1}^{n_{2}}&\frac{\Gamma(q^{-\nu_{2,i}}t^{-n_{2}+i})\Gamma((f_{1}^{+}/f_{2}^{+})\,q^{-\nu_{2,i}}t^{-n_{2}+i})}{\Gamma(v^{2}(f_{1}^{-}/f_{2}^{+})\,q^{-\nu_{2,i}}t^{-n_{2}+i})\Gamma(v^{2}(f_{2}^{-}/f_{2}^{+})\,q^{-\nu_{2,i}}t^{-n_{2}+i})}\\
\times&\frac{\Gamma(v^{2} (f_{1}^{-}/f_{2}^{+})\,t^{-n_{2}+i})\Gamma(v^{2} (f_{2}^{-}/f_{2}^{+})\,t^{-n_{2}+i})}{\Gamma(t^{-n_{2}+i})\Gamma((f_{1}^{+}/f_{2}^{+})\,t^{-n_{2}+i})}.
\end{align} 
\subsection{6d $U(3)$ Theory with $N_{f}=6$ }
\par{In the same spirit as in \cite{Aganagic:2013tta}, the Dotsenko-Fateev representation of the conformal blocks for the 5 point function does not reproduce the instanton partition function of the gauge theory according to the AGT conjecture, but instead its spectral/fiber-base dual. Therefore, we need to compute the instanton partition function of the $U(3)$ theory with $N_{f}=6$. The refined vertex computation can be generalized for this case using curve counting arguments. The partition function becomes is given by}
\begin{align}\nonumber
Z&=\sum_{\N,\NN,\NNN}(vQ_{b_{1}})^{|\N|}(vQ_{b_{2}})^{|\NN|}(vQ_{b_{3}})^{|\NNN|}\prod_{k=1}^{\infty}N_{\N\es}(v\qt^{k-1}\,e_{1}/f_{1}^{-})N_{\es\N}(v\qt^{k}\,f_{1}^{-}/e_{1})\\\nonumber&\times N_{\NN\es}(v\qt^{k-1}\,e_{2}/f_{1}^{-})N_{\es\NN}(v\qt^{k}\,f_{1}^{-}/e_{2})N_{\NNN\es}(v\qt^{k-1}\,e_{3}/f_{1}^{-})N_{\es\NNN}(v\qt^{k}\,f_{1}^{-}/e_{3})\\\nonumber&\times N_{\es\N}(v\qt^{k-1}\,f_{2}^{-}/e_{1})N_{\N\es}(v\qt^{k}\,e_{1}/f_{2}^{-})N_{\NN\es}(v\qt^{k-1}\,e_{2}/f_{2}^{-})N_{\es\NN}(v\qt^{k}\,f_{2}^{-}/e_{2})\\\nonumber&\times N_{\NNN\es}(v\qt^{k-1}\,e_{3}/f_{2}^{-})N_{\es\NNN}(v\qt^{k}\,f_{2}^{-}/e_{3})N_{\es\N}(v\qt^{k-1}\,f_{3}^{-}/e_{1})N_{\N\es}(v\qt^{k}\,e_{1}/f_{3}^{-})\\\nonumber &\times N_{\es\NN}(v\qt^{k-1}\,f_{3}^{-}/e_{2})N_{\NN\es}(v\qt^{k}\,e_{2}/f_{3}^{-})N_{\NNN\es}(v\qt^{k-1}\,e_{3}/f_{3}^{-})N_{\es\NNN}(v\qt^{k}\,f_{3}^{-}/e_{3})\\\nonumber &\times N_{\es\N}(v\qt^{k-1}\,f_{1}^{+}/e_{1})N_{\N\es}(v\qt^{k}\,e_{1}/f_{1}^{+})N_{\es\N}(v\qt^{k-1}\,f_{2}^{+}/e_{1})N_{\N\es}(v\qt^{k}\,e_{1}/f_{2}^{+})\\\nonumber & \times N_{\es\N}(v\qt^{k-1}\,f_{3}^{+}/e_{1})N_{\N\es}(v\qt^{k}\,e_{1}/f_{3}^{+})N_{\NN\es}(v\qt^{k-1}\,e_{2}/f_{1}^{+})N_{\es\NN}(v\qt^{k}\,f_{1}^{+}/e_{2})\\\nonumber & \times N_{\es\NN}(v\qt^{k-1}\,f_{2}^{+}/e_{2})N_{\NN\es}(v\qt^{k}\,e_{2}/f_{2}^{+})N_{\es\NN}(v\qt^{k-1}\,f_{3}^{+}/e_{2})N_{\NN\es}(v\qt^{k}\,e_{2}/f_{3}^{+})\\\nonumber & \times N_{\NNN\es}(v\qt^{k-1}\,e_{3}/f_{1}^{+})N_{\es\NNN}(v\qt^{k}\,f_{1}^{+}/e_{3})N_{\NNN\es}(v\qt^{k-1}\,e_{3}/f_{2}^{+})N_{\es\NNN}(v\qt^{k}\,f_{2}^{+}/e_{3})\\\nonumber & \times N_{\es\NNN}(v\qt^{k-1}\,f_{3}^{+}/e_{3})N_{\NNN\es}(v\qt^{k}\,e_{3}/f_{3}^{+}) N_{\NN \N}^{-1}(v^2\qt^{k-1}\,e_{2}/e_{1})N_{\N \NN}^{-1}(v^2\qt^{k}\,e_{1}/e_{2}) \\\nonumber & \times            N_{\NNN\N}^{-1}(v^2\qt^{k-1}\,e_{3}/e_{1})N_{\N\NNN}^{-1}(v^2\qt^{k}\,e_{1}/e_{3})N_{\NNN\NN}^{-1}(v^2\qt^{k-1}\,e_{3}/e_{2})N_{\NN\NNN}^{-1}(v^2\qt^{k}\,e_{2}/e_{3})\\\nonumber &\times N_{\N\N}^{-1}(v^2\qt^{k-1})N_{\NN\NN}^{-1}(v^2\qt^{k-1})N_{\NNN\NNN}^{-1}(v^2\qt^{k-1})N_{\NN\N}^{-1}(\qt^{k-1}\,e_{2}/e_{1})N_{\N\NN}^{-1}(\qt^{k}\,e_{1}/e_{2})\\\nonumber & \times N_{\NNN\N}^{-1}(\qt^{k-1}\,e_{3}/e_{1}) N_{\N\NNN}^{-1}(\qt^{k}\,e_{1}/e_{3})N_{\NNN\NN}^{-1}(\qt^{k-1}\,e_{3}/e_{2})N_{\NN\NNN}^{-1}(\qt^{k}\,e_{2}/e_{3}) N_{\N\N}^{-1}(\qt^{k})\\ &\times N_{\NN\NN}^{-1}(\qt^{k})N_{\NNN\NNN}^{-1}(\qt^{k})
\end{align}
After relatively tedious algebra we recognize that the elliptic Vandermonde is produced and the potentials take the generic form for $M$ fundamental hypermultiplets
\begin{align}\nonumber
\frac{\left[\prod_{a=1}^{M}V_{a}(y)\right]_{y=y_{\nu_{a}}}}{\left[\prod_{a=1}^{M}V_{a}(y)\right]_{y=y_{\es}}}=\prod_{1\leq a,b\leq M}\prod_{i=1}^{n_{a}}&\frac{\Gamma((f^{+}_{b}/f^{+}_{a})q^{-\nu_{a,i}}t^{-n_{a}+i})\Gamma(v^{2}(f^{-}_{b}/f^{+}_{a})t^{-n_{a}+i})}{\Gamma(v^2(f_{b}^{-}/f_{a}^{+})\,q^{-\nu_{a,i}}t^{-n_{a}+i})\Gamma((f^{+}_{b}/f^{+}_{a})t^{-n_{a}+i})},
\end{align} 
those generalizing to an arbitrary linear conformal block.
\section{Discussion}
In this paper we have established a connection between elliptic deformation of vertex operators that appear in the calculation of topological string partition functions via the topological vertex and compactification of web diagrams corresponding to Calabi-Yau threefolds. Using this connection we saw that the elliptic deformation of the conformal blocks is related to the 6d gauge theories that arise from Calabi-Yau threefolds dual to the compactified brane webs.

It will be interesting to study the worldsheet description of this elliptic deformation as we have only studied the ellipticization of the corresponding vertex operators. Similarly it would be useful to study the toric geometry associated with compactified Newton polygons. Some progress in this direction has been made \cite{DK} and it would useful to spell out the details for all webs which can be compactified.

\section*{Acknowledgement}
We would like to thank Giulio Bonelli, Nathan Haouzi, Shamil Shakirov, Dan Xie, Cumrun Vafa, Yegor Zenkevich for many valuable discussions. CK would like to thank the organizers and the participants of the ``Workshop on Geometric Correspondences of Gauge Theories'' at SISSA. This work was  supported by the Center for Mathematical Sciences and Applications at Harvard University.


\section{Appendix}
In this appendix, we would like to collect some definitions, identities to fix our conventions and some details of the computations. The instanton partition function is computed by the refined topological vertex \cite{Iqbal:2007ii}; it can be expressed in terms of Macdonald function and skew Schur fucntions:
\begin{align}
C_{\lambda\,\mu\,\nu}(t,q)=t^{-\frac{\Arrowvert\mu^t\Arrowvert^2}{2}}\,q^{\frac{\Arrowvert\mu\Arrowvert^2+\Arrowvert\nu\Arrowvert^2}{2}}\,\widetilde{Z}_{\nu}(t,q)
\sum_{\eta}\Big(\frac{q}{t}\Big)^{\frac{|\eta|+|\lambda|-|\mu|}{2}}\,s_{\lambda^{t}/\eta}(t^{-\rho}\,q^{-\nu})\,s_{\mu/\eta}(t^{-\nu^t}\,q^{-\rho})\,,
\end{align}
where $\rho=\{-\frac{1}{2},-\frac{3}{2},-\frac{5}{2},\mathellipsis\}$ is the Weyl vector for $SU(\infty)$, and $s_{\nu}(x_{1},x_{2},\mathellipsis)$ is the Schur function labelled by a partition $\nu$, and
\begin{align}
\widetilde{Z}_{\nu}(t,q)=\prod_{(i,j)\in \nu}\Big(1-q^{\nu_{i}-j}\,t^{\nu^{t}_{j}-i+1}\Big)^{-1},
\end{align}
and is proportional to Macdonald functions at the \textit{special point} $x_{i}=t^{i-1/2}$,
\begin{align}
P_{\nu}(t^{-\rho};q,t)=t^{\frac{\Arrowvert\nu^t\Arrowvert^2}{2}}\,\widetilde{Z}_{\nu}(t,q).
\end{align} 
We have heavily relied on the following identities 
\begin{align}
&n(\lambda)\equiv\sum_{i=1}^{\ell(\lambda)}(i-1)\lambda_{i}=\frac{1}{2}\sum_{i=1}^{\ell(\lambda)}\lambda^{t}_{i}(\lambda^{t}_{i}-1)=\sum_{(i,j)\in\lambda}(\lambda_{j}^{t}-i)=\frac{\Arrowvert\lambda^{t}\Arrowvert^{2}}{2}-\frac{|\lambda|}{2}\,,\\\label{n2}
&n(\lambda^{t})\equiv\sum_{i=1}^{\ell(\lambda^{t})}(i-1)\lambda^{t}_{i}=\frac{1}{2}\sum_{i=1}^{\ell(\lambda^{t})}\lambda_{i}(\lambda_{i}-1)=\sum_{(i,j)\in\lambda}(\lambda_{i}-j)=\frac{\Arrowvert\lambda\Arrowvert^{2}}{2}-\frac{|\lambda|}{2}\,,
\end{align}
where $\Arrowvert\mu\Arrowvert^{2}=\sum_{i=1}^{\ell(\mu)}\mu_{i}^2$, $\ell(\mu)$ is denoting the length of Young diagram $\mu$, in other words the number of non-zero $\mu_{i}$'s. Furthermore, we can define the hook length $h(i,j)$ and the content $c(i,j)$ of a given box in a particular Young diagram 
\begin{align}
h(i,j)=\mu_{i}-j+\mu_{j}^{t}-i+1,\qquad c(i,j)=j-i\,,
\end{align}
which satisfy
\begin{align}
&\sum_{(i,j)\in\mu}h(i,j)=n(\mu)+n(\mu)+|\mu|,\\
&\sum_{(i,j)\in\mu}c(i,j)=n(\mu^{t})-n(\mu)=\frac{1}{2}\Arrowvert\mu\Arrowvert^{2}-\frac{1}{2}\Arrowvert\mu^{t}\Arrowvert^{2}\equiv \frac{1}{2}\kappa(\mu).
\end{align}
The following sum rules are essential for vertex computations \cite{macdonald}
\begin{align}
&\sum_{\eta}s_{\eta/\lambda}(\mathbf{x})s_{\eta/\mu}(\mathbf{y})=\prod_{i,j=1}^{\infty}(1-x_{i}y_{j})^{-1}\sum_{\tau}s_{\mu/\tau}(\mathbf{x})s_{\lambda/\tau}(\mathbf{y})\,.\\
&\sum_{\eta}s_{\eta^{t}/\lambda}(\mathbf{x})s_{\eta/\mu}(\mathbf{y})=\prod_{i,j=1}^{\infty}(1+x_{i}y_{j})\sum_{\tau}s_{\mu^{t}/\tau}(\mathbf{x})s_{\lambda^{t}/\tau^{t}}(\mathbf{y})\,.
\end{align}
The instanton partition function is given by the normalized topological string partition function. After cutting the toric geometry into the two so-called strip geometries, we compute the open topological string amplitude with the boundary condition labelled by Young diagrams. The instanton partition function is this open amplitude divided by the closed amplitude. The following identity is essential to get the instanton partition function:

\begin{align}
\prod_{i,j=1}^{\infty}\frac{1-Q\,q^{\nu_{i}-j}t^{\mu_{j}^{t}-i+1}}{1-Q\,q^{-j}t^{-i+1}}=\prod_{(i,j)\in\nu}\Big(1-Q\,q^{\nu_{i}-j}t^{\mu_{j}^{t}-i+1}\Big)\prod_{(i,j)\in\mu}\Big(1-Q\,q^{-\mu_{i}+j-1}t^{-\nu_{j}^{t}+i}\Big),
\end{align}

The (first) $\theta$-function and the Dedekind $\eta$-function are defined by
\bea\nn
\theta_{1}(\tau;z)&=&-i\,e^{\frac{i\pi\,\tau}{4}}\,e^{i\pi z}\prod_{k=1}^{\infty}\Big[(1-e^{2\pi\,i\,k\tau})(1-e^{2\pi\,i\,k\tau}\,e^{2\pi i\,z})(1-e^{2\pi\,i\,(k-1)\tau}\,e^{-2\pi i\,z})\Big]\\\nn
&=&-i\,e^{\frac{i\pi\,\tau}{4}}\,(e^{i\pi z}-e^{-i\pi z})\prod_{k=1}^{\infty}\Big[(1-e^{2\pi\,i\,k\tau})(1-e^{2\pi\,i\,k\tau}\,e^{2\pi i\,z})(1-e^{2\pi\,i\,k\tau}\,e^{-2\pi i\,z})\Big]\\\nn
\eta(\tau)&=&e^{\frac{i\pi \tau}{12}}\,\prod_{k\geq 1}\Big(1-e^{2\pi i\,k\tau}\Big)
\eea



\end{document}